\newif\iflongversion
\longversiontrue

\documentclass[USenglish,oneside,twocolumn]{article}

\usepackage[utf8]{inputenc}\usepackage{xspace}
\iflongversion
\else
\usepackage[big]{dgruyter_NEW}
\fi
\usepackage{color}
\usepackage{latexsym}
\usepackage{hyperref}
\usepackage{multirow}
\usepackage{comment}
\usepackage{enumerate}
\usepackage{graphicx}
\usepackage{tikz}
\usepackage{wasysym}
\usepackage{xcolor}
\usepackage{algorithm}
\usepackage{algpseudocode}
\usepackage{fullwidth}
\usepackage{mdwlist}
\usepackage{booktabs}
\usepackage{balance}
\usepackage{listings}
\usepackage{bm}                 \usepackage{lmodern}
\usepackage[normalem]{ulem}
\usepackage{moresize}
\usepackage{pdfpages}
\usepackage{adjustbox}
\usepackage{array}

\newcommand{\mytitle}{Holes in the Geofence: Privacy Vulnerabilities in ``Smart''~DNS~Services}

\hypersetup{
     bookmarks=true,
     colorlinks=true,
     linkcolor=black,
     citecolor=black,
     filecolor=black,
     urlcolor=black,
     pdfauthor = {Hidden for double-blind review},
     pdftitle = {\mytitle},
     pdfkeywords = {}
}

\PassOptionsToPackage{hyphens}{url}
\makeatletter
\g@addto@macro{\UrlBreaks}{\UrlOrds}
\makeatother

\newcommand*\circled[1]{{\small \tikz[baseline=(char.base)]{
            \node[shape=circle,fill,inner sep=.5pt] (char)
            {\textcolor{white}{{\bf #1}}};}}}

\algnewcommand{\LineComment}[1]{\State \(\triangleright\) #1}

\newcommand\pagebudget[1]{}

\newcommand{\threatmodel}[1]{\noindent \begin{small}{\color{red} \fbox{{\sf Threat model(s): #1}}}\end{small}\smallskip\noindent}

\newcommand{\Section}[1]{\vspace{-.20in}\section{#1}\vspace{-.1in}}
\newcommand{\Subsection}[1]{\vspace{-.15in}\subsection{#1}\vspace{-.1in}}
  
\definecolor{Red}{rgb}{1,0,0}
\definecolor{Orange}{rgb}{1,0.5,0}
\definecolor{Green}{rgb}{0,1,0}
\definecolor{Blue}{rgb}{0,0,1}
\definecolor{Magenta}{rgb}{1,0,1}
\definecolor{DarkGreen}{rgb}{0.0, 0.5, 0.0}

\newcommand\breakme[0]{}

\newcommand\added[1]{\textcolor{black}{{#1}}}

\newcommand\confirmed[1]{#1}

\newcommand\changed[1]{#1}

\ifdefined\Paragraph
\renewcommand\Paragraph[1]{
  \vspace{.05in}
  \noindent {\bf #1.}\quad }
\else
\newcommand\Paragraph[1]{
  \vspace{.05in}
  \noindent {\bf #1.}\quad }
\fi

\newcommand{\rahel}[1]{\breakme{\noindent\color{violet}\em [Rahel: #1]}}
\newcommand{\adam}[1]{{\noindent\color{Magenta}\em [Adam: #1]}}

\DeclareTextFontCommand{\mytexttt}{\ttfamily\hyphenchar\font=45\relax}
\newcommand{\htmlize}[1]{{\small \mytexttt{#1}}\xspace}

\newcommand{\sdns}[0]{SDNS\xspace}

\iflongversion
\else
\cclogo{\includegraphics{by-nc-nd.pdf}}
\fi

\begin{document}

\iflongversion
\title{\mytitle\thanks{A shorter version of this paper appears in {\bf
      Proceedings on Privacy Enhancing Technologies (PoPETS)}, July 2021.}}
\author{Rahel A. Fainchtein$^*$ \and
Adam J. Aviv$^+$ \and
Micah Sherr$^*$ \and
Stephen Ribaudo$^*$ \and
Armaan Khullar$^*$
}
\date{$^*$ Georgetown University \quad $^+$ The George Washington University}
\fi

\maketitle

\iflongversion
\begin{abstract}
{{\em Smart DNS} (\sdns) services advertise access to {\em geofenced} content (typically, video streaming sites such as Netflix or Hulu) that is normally inaccessible unless the client is within a prescribed geographic region.  
	\sdns is simple to use and involves no software installation. Instead, it requires only that users modify their DNS settings to point to an \sdns resolver. 
	The \sdns resolver ``smartly'' identifies geofenced domains and, in lieu of their proper DNS resolutions, returns IP addresses of proxy servers located within the geofence.  
	These servers then transparently proxy traffic between the users and their intended destinations, allowing for the bypass of these geographic restrictions.\\ 
  \indent\quad This paper presents the first academic study of \sdns services.  We identify a number of serious and pervasive privacy vulnerabilities that expose information about the users of these systems. 
  These include architectural weaknesses that enable content providers to identify which requesting clients use \sdns.
  Worse, we identify flaws in the design of some \sdns services that allow {\em any} arbitrary third party to enumerate these services' users (by IP address), even if said users are currently offline.  We present mitigation strategies to these attacks that have been adopted by at least one \sdns provider in response to our findings.
}
  
  \end{abstract}

 \fi

\newcommand{\mylabel}[1]{~\\{\centerline {\bf #1}}}

\newcolumntype{R}[2]{>{\adjustbox{angle=#1,lap=\width-(#2)}\bgroup}l<{\egroup}}
\newcommand*\rot{\multicolumn{1}{R{45}{1em}}}

\newcommand{\tabpopularity}[0]{
\begin{table*}[t]
  \caption{Derived most popular channels and average estimated resolution rate
    ($\lambda$), in requests per hour.}
  \label{tbl:channelpopularity}
  \begin{minipage}[t]{.22\linewidth}
    \begin{scriptsize}
      \resizebox{\linewidth}{!}{\begin{tabular}{p{.87in}p{.8in}}
       {\bf Site} & {\bf $\bm{\lambda}$ (95\% CI)} \\
       \toprule
       tvland.com & 47526 $\pm$65 \\
       player.pl & 47462 $\pm$131 \\
       hbogo.com & 47413 $\pm$76 \\
       pandora.com & 47339 $\pm$93 \\
       comedycentral.com & 47339 $\pm$158 \\
       sonycrackle.com & 39475 $\pm$2385 \\
       theloop.ca & 39099 $\pm$4992 \\
       absoluteradio.co.uk & 38714 $\pm$217K \\
       amazon.co.uk & 36712 $\pm$2224 \\
       bleacherreport.com & 30703 $\pm$6217 \\
       \bottomrule
     \end{tabular}} \\
    \mylabel{CactusVPN}
    \end{scriptsize}
  \end{minipage}
  \hfill
  \begin{minipage}[t]{.22\linewidth}
    \begin{scriptsize}
      \resizebox{\linewidth}{!}{\begin{tabular}{p{.87in}p{.8in}}
    	{\bf Site} & {\bf $\bm{\lambda}$ (95\% CI)} \\
    	\toprule
    	amazon.com & 95694 $\pm$401509 \\
    	youtube.com & 79822 $\pm$79257 \\
    	foxsportsgo.com & 70913 $\pm$5113 \\
    	bloomberg.com & 46726 $\pm$5174 \\
    	disneylife.com & 46498 $\pm$673 \\
    	funimation.com & 46342 $\pm$247 \\
    	theloop.ca & 46122 $\pm$278 \\
    	travelchannel.com & 46036 $\pm$874 \\
    	amctv.com & 45987 $\pm$44 \\
    	viaplay.se & 45948 $\pm$39 \\
    	\bottomrule
    \end{tabular}}\\
    \mylabel{SmartDNSProxy}
    \end{scriptsize}
  \end{minipage}
      \hfill
  \begin{minipage}[t]{.22\linewidth}
    \begin{scriptsize}
      \resizebox{\linewidth}{!}{\begin{tabular}{p{.87in}p{.8in}}
        {\bf Site} & {\bf $\bm{\lambda}$ (95\% CI)} \\
        \toprule
        foxsports.com.au & 1.7M $\pm$2.7M \\ zdf.de & 395633 $\pm$2482 \\
        m6replay.fr & 393483 $\pm$2741 \\
        tsn.ca & 91665 $\pm$1662 \\
        rds.ca & 88007 $\pm$1163 \\
        tennischannel... & 72537 $\pm$2066 \\
        hbonow.com & 66332 $\pm$14379 \\
        itv.co & 59623 $\pm$15541 \\
        itv.com & 44405 $\pm$872 \\
        foxsoccer2go.com & 34031 $\pm$209 \\
        \bottomrule
      \end{tabular}} \\
      \mylabel{IB VPN}
    \end{scriptsize}
  \end{minipage}
  \hfill
  \begin{minipage}[t]{.22\linewidth}
    \begin{scriptsize}
      \resizebox{\linewidth}{!}{\begin{tabular}{p{.87in}p{.8in}}
        {\bf Site} & {\bf $\bm{\lambda}$ (95\% CI)} \\
        \toprule
        amazon.co.uk & 754402 $\pm$108K \\
        oxygen.com & 497202 $\pm$345K \\
        magine.com & 341406 $\pm$468 \\
        songza.com & 324251 $\pm$4487 \\
        vtele.ca & 318096 $\pm$4708 \\
        abema.tv & 285243 $\pm$3549 \\
        cbc.ca & 285113 $\pm$307K \\
        telemundo.com & 271029 $\pm$211K \\
        rte.ie & 268518 $\pm$16790 \\
        tennischannel.com & 261632 $\pm$15117 \\
        \bottomrule
      \end{tabular}}\\
      \mylabel{IronSocket}
    \end{scriptsize}
  \end{minipage}\\
  \hfill
  \begin{minipage}[t]{.22\linewidth}
    \begin{scriptsize}
      \resizebox{\linewidth}{!}{\begin{tabular}{p{.87in}p{.8in}}
        {\bf Site} & {\bf $\bm{\lambda}$ (95\% CI)} \\
        \toprule
        starzplay.com & 90721 $\pm$43758 \\
        cartoonnetwork.com & 88345 $\pm$216417 \\
        ahctv.com & 87648 $\pm$91137 \\
        cwtv.com & 79143 $\pm$74440 \\
        cwseed.com & 78446 $\pm$51686 \\
        indieflix.com & 71231 $\pm$219415 \\
        theonion.com & 35655 $\pm$11610 \\
        teamcoco.com & 35201 $\pm$137079 \\
        tlc.com & 35200 $\pm$1.52M \\
        showtime.com & 35196 $\pm$118214 \\
        \bottomrule
      \end{tabular}}\\
      \mylabel{Keenow}
    \end{scriptsize}
  \end{minipage}
 \hfill
 \begin{minipage}[t]{.22\linewidth}
    \begin{scriptsize}
      \resizebox{\linewidth}{!}{\begin{tabular}{p{.87in}p{.8in}}
    		{\bf Site} & {\bf $\bm{\lambda}$ (95\% CI)} \\
    		\toprule
    		instagram.com & 2708804 $\pm$97K \\
    		epix.com & 116632 $\pm$388 \\
    		vh1.com & 116564 $\pm$7429 \\
    		discovery.com & 113888 $\pm$21888 \\
    		cnbc.com & 113722 $\pm$9419 \\
    		history.com & 111340 $\pm$2690 \\
    		amc.com & 109985 $\pm$2764 \\
    		aetv.com & 108331 $\pm$1187 \\
    		beinsports.com & 106153 $\pm$1044 \\
    		cnn.com & 101036 $\pm$187K \\
    		\bottomrule
    	\end{tabular}} \\
    	\mylabel{DNSTrick}
    \end{scriptsize}
  \end{minipage}
  \hfill
  \begin{minipage}[t]{.22\linewidth}
    \begin{scriptsize}
      \resizebox{\linewidth}{!}{\begin{tabular}{p{.87in}p{.80in}}
        {\bf Site} & {\bf $\bm{\lambda}$ (95\% CI)} \\
        \toprule
        player.pl & 423944 $\pm$829 \\
        hbogo.com & 421225 $\pm$683 \\
        pandora.com & 413256 $\pm$1182 \\
        tvland.com & 406346 $\pm$4193 \\
        comedycentral.com & 404203 $\pm$5509 \\
        amazon.co.uk & 306602 $\pm$209K \\
        cnbc.com & 252442 $\pm$593K \\
        sling.com & 208366 $\pm$34462 \\
        nickjr.com & 202391 $\pm$46491 \\
        foxsports.com & 197627 $\pm$20198 \\
        \bottomrule
      \end{tabular}}\\
      \mylabel{SmartyDNS}
    \end{scriptsize}
  \end{minipage}
 \hfill
  \begin{minipage}[t]{.22\linewidth}
    \begin{scriptsize}
      \resizebox{\linewidth}{!}{\begin{tabular}{p{.87in}p{.80in}}
        {\bf Site} & {\bf $\bm{\lambda}$ (95\% CI)} \\
        \toprule
        docclub.com & 79823 $\pm$67529 \\
        discovery.com & 78740 $\pm$48817 \\
        showcase.ca & 74609 $\pm$58135 \\
        tvplayer.com & 66103 $\pm$90796 \\
        zattoo.com & 59129 $\pm$54614 \\
        syfy.com & 34851 $\pm$31357 \\
        nbc.com & 33614 $\pm$6276 \\
        history.com & 33509 $\pm$2476 \\
        rdio.com & 33331 $\pm$3398 \\
        klowdtv.com & 33275 $\pm$67988 \\
        \bottomrule
      \end{tabular}}\\
      \mylabel{TrickByte}
    \end{scriptsize}
  \end{minipage}
\end{table*}}

\newcommand{\tabrevenue}[0]{
  \begin{table}[t]
  \caption{The estimated number of users of a single \sdns resolver for various \sdns
    providers, and the estimated monthly profit.}
  \label{tbl:estusers}
  \begin{center}
  \begin{small}
    \begin{tabular}{lrrr}
      {\bf Service} & {\bf Rate} $\bm{(\lambda)}$ & {\bf Est.~Users ($\bm{n}$)} & {\bf Est.~Profit} \\ \toprule
      CactusVPN & 41\,,119 & 15\,,635 & \$ 76\,,977\\
      DNStrick & 1\,,794 & 682  & \$ 3\,,330\\
      HideIP VPN & 2\,,127 & 809  & \$ 3\,,952\\
      SmartyDNS & 6\,,389 & 2\,,429 & \$ 11\,,741\\
      TrickByte & 8\,,269 & 3\,,144  & \$ 9\,,190\\
      Unlocator & 3\,,565 & 1\,,356 & \$ 6\,,622\\
      \bottomrule
    \end{tabular}\\
  \end{small}
  \end{center}
\end{table}}

\newcommand{\tablecounts}[0]{

  \begin{table*}[t]
  \centering
  
  \begin{minipage}{.48\linewidth}

    \caption{Top 10 Countries with the most \sdns resolvers}
    \label{tab:resolver:loc}
    \centering
    \begin{tabular}{r l}
      {\bf  Country} & {\bf Num. Resolvers} \\ \hline
      \toprule
      United States  & 9\\
      United Kingdom & 4 \\
      Canada & 4  \\
      Australia & 3 \\
      Germany & 3 \\
      India & 3 \\
      Netherlands & 3 \\
      Denmark & 2 \\
      South Africa & 2  \\
      Singapore & 2 \\
      \bottomrule
    \end{tabular}

 \vspace{2em}

 \caption{Top ASes with the most \sdns resolvers}
 \label{tab:resolver:AS}
 \begin{tabular}{r l}
   {\bf AS Name and Number} & {\bf Num. Resolvers} \\ \hline			
   \toprule
   Amazon (16509) & 21\\
   DigitalOcean (14061) & 15 \\
   SoftLayer Technologies (36351) & 10 \\
   Choopa LLC  (20473) & 5 \\
   Iomart (20860) & 5\\
   Linode LLC (63949) & 3\\
   OVH SAS (16276) & 3\\
   SiteHost New Zealand (45179) & 3\\
   ASERGO Scandinavia ApS (30736)  & 2\\
   Datacamp Ltd (60068) & 2\\
   \bottomrule
 \end{tabular}
  \end{minipage}
\hfill
\begin{minipage}{.48\linewidth}
    \centering
    \caption{Top 10 Countries with the most proxies.}
    \label{topCountries}
    \label{tab:proxy:loc}
      \begin{tabular}{r l}
        {\bf  Country} & {\bf Num. Proxies} \\ \hline
        \toprule
        United States  & 15\\
        United Kingdom & 13 \\
        India & 6 \\
        Australia & 3 \\
        Denmark & 3 \\
        Sweden & 2  \\
        France & 2 \\
        Canada & 2  \\
        Germany & 2 \\
        Norway & 1 \\
        \bottomrule
      \end{tabular}

      \vspace{2.5em}
     \caption{Top ASes with the most proxies.}
     \label{topAS-es}
     \label{tab:proxy:AS}
     \centering
       \begin{tabular}{r l}
         {\bf AS Name and Number} & {\bf Num. Proxies} \\ \hline			
         \toprule
         DigitalOcean (14061) & 8\\
         QuadraNet (8100) & 6 \\
         Iomart (20860) & 4 \\
          Level 3 Parent LLC  (3356) & 4  \\
          ASERGO Scandinavia ApS (30736)  & 3\\
         OVH SAS (16276) & 2\\
         GleSYS AB (42708)  & 2\\
         Compuweb (51905) & 2 \\
         Melbikomas UAB (56630) & 1 \\
         \bottomrule
       \end{tabular}
       \vspace{+.2in}
   \end{minipage}
\end{table*}}

\newcommand{\tabproviders}[0]{
  \begin{table}[t]
  \caption{Identified \sdns providers. As indicated by their names, many double as purveyors of commercial VPN access.}
  \label{tbl:costs}
  \label{tab:providers}
  \centering
    \begin{tabular}{rll}
      {\bf Provider} & {\bf Monthly Cost} & {\bf Location} \\
      \toprule
      AceVPN & \$\,5.95\,$^{\circ}$ &  USA \\ 
      Blockless & \$\,3.32\,$^{\circ}$ & Canada  \\ 
      ${}^\star$BulletVPN & \$\,10.98$^{\circ}$ & Estonia  \\ 
      ${}^\star$CactusVPN & \$\,4.99 & Moldova  \\
      ${}^\star$DNSFlex & \$\,5.00 & Canada \\
      ${}^\star$DNSTrick & \$\,4.95 & Unknown \\
      ${}^\star$GetFlix & \$\,39.00$^{\bullet}$  & Turkey \\
      ${}^\star$HideIPVPN & \$\,4.95 & Unknown  \\
      Hide-my-IP & \$\,4.95 & USA \\
${}^\star$ibVPN & \$\,10.95 & Romania  \\
      Ironsocket & \$\,4.16 & Hong Kong \\
      ${}^\star$Keenow & \$\,5.79 & Israel  \\
      Le-VPN & \$\,9.95$^{\circ}$ & Hong Kong  \\ 
      ${}^\star$Overplay & \$\,4.99 & USA \\
      simpletelly  & \$\,4.99 & Turkey \\
      ${}^\star$SmartDNSproxy & \$\,4.90 & Turkey  \\
      ${}^\star$SmartyDNS & \$\,4.90 & Moldova  \\
      StrongDNS & \$\,5.00 & USA\\
      ${}^\star$TrickByte & \$\,2.99 & Turkey \\
      TVWhenAway & \pounds\,7.99 & UK \\
      ${}^\star$Uflix& \$\,4.90$^{\circ}$ & Turkey  \\ 
      Unblock-us & \$\,4.99 & Cyprus \\
      ${}^\star$ Unlocator & \$\,4.95 & Denmark  \\ 
      VPNSecure & \$\,9.95 & Australia  \\
      ${}^\star$VPNUK & \pounds\,5.99 & UK${}^\diamond$  \\      
      \midrule
      \multicolumn{3}{l}{${}^\star$ indicates a provider included in our measurement analysis} \\
      \multicolumn{3}{l}{${}^\bullet$ indicates a lifetime cost} \\
      \multicolumn{3}{l}{${}^{\circ}$ indicates a cost that also includes VPN services}\\
      \multicolumn{3}{l}{${}^\diamond$ contact info in UK, but company registered in Belize} \\
      \bottomrule      
    \end{tabular}
\end{table}}

\newcommand{\tableproxyuserauth}[0]{
\begin{table}[t]
	\caption{Occurence of open and universal proxies, by \sdns
          provider, for both HTTP and SNI proxy methods.   Uflix, Trickbyte, and SmartDNSProxy shared
          10 proxy servers; SmartDNSProxy and Trickbyte shared an
          additional seven proxies; and CactusVPN and SmartyDNS used
          the same  five proxies. Moreover, among the \sdns providers studied, 
          we  observed that, for each protocol supported, an \sdns provider's proxies either were all open/universal, or none of them were. }
        \label{tab:ouproxies}
\small
\begin{tabular}{r c c c c c}
          {\bf Provider} &  \rot{\bf Confirmed Proxies} &
              \rot{\bf Open (\htmlize{HOST}/HTTP)} &
              \rot{\bf Universal (\htmlize{Host}/HTTP)} &  
              \rot{\bf Open (SNI/HTTPS)} & 
              \rot{\bf Universal (SNI/HTTPS)} \\
                                                        \toprule

    	CactusVPN & 5 & \Circle & \CIRCLE & \CIRCLE & \CIRCLE \\          
       HideIPVPN & 3 & \Circle & \CIRCLE & \CIRCLE & \CIRCLE \\ 
    	IBVPN & 1 & \Circle & \CIRCLE & \Circle & \CIRCLE \\
        SmartDNSProxy & 36 &\Circle & \Circle & \Circle & \Circle \\
    	SmartyDNS & 5 & \Circle & \CIRCLE & \CIRCLE & \CIRCLE \\          
    	Trickbyte & 18& \Circle & \Circle & \Circle & \Circle \\
    	Uflix & 14 & \Circle & \Circle & \Circle & \Circle\\
    	VPNUK & 17 & \Circle & \CIRCLE & \Circle & \CIRCLE \\
          \toprule
        \end{tabular}
\begin{minipage}{\linewidth}
          \medskip
          \small
          \Circle \quad none of the service's tested proxies operated in this mode\\
          \CIRCLE \quad all of the service's tested  proxies operated
          in this mode \\
\end{minipage}
\end{table}}

\newcommand{\tabletraceroute}[0]{
  \begin{table}[t]
    \caption{\changed{Average number of ASes encountered in network paths
        from various geographic regions to (1)~Cloudflare's and
        Google's DNS resolvers (``Public Resolver'')  and (2)~108 \sdns
        resolvers (``SDNS Resolver''). Percentage increases (relative to the public
        resolvers) are shown in parentheses.}}
    \label{tbl:traceroute}
    \small
    \centering
    \begin{tabular}{c c c}
      {\bf Client Location} & {\bf Public Resolver} & {\bf \sdns Resolver}
      \\
      \toprule
      Australia & 1.50 & 2.74 (82.67\%$\uparrow$) \\
      Belgium & 1.00 & 2.60 (160.00\%$\uparrow$)\\
      Brazil & 2.00 & 2.31 (15.50\%$\uparrow$) \\
      Japan & 2.00 & 2.26 (13.00\%$\uparrow$) \\
      United States & 2.00 & 3.10 (55.00\%$\uparrow$) \\
      \bottomrule
    \end{tabular}
\end{table}
}

\newcommand{\tablethreats}[0]{
  \begin{table*}[t]
  \caption{\changed{Summary of attacks.  The adversary, required
      adversary capabilities, and the target of the attack are listed
      for each attack.}}
  \label{tbl:threats}
  \small
  \centering
  \begin{tabular}{c c c c}
    {\bf Vulnerability} & {\bf Adversary} & {\bf Required Adversary
                                            Cap.} & {\bf Target}
    \\
    \toprule
      \S\ref{sec:enumeration}: Enumerating customers (by IP) & Internet user & reg.~domain name;
                                                    spoof UDP &
                                         Customer Threat
                                                         
    \\

\S\ref{sec:deproxying}:    Real-time \sdns customer identification                 &
    Content provider & operate website; view web logs & Customer Threat 
                                                           
    \\
                   
                                                           \S\ref{sec:deproxying}:        Real-time proxy server discovery                 &
    Content provider & operate website; view web logs & \sdns Provider Threat
    \\

   \S\ref{sec:eavesdropping}:    Increased risk of traffic analysis &
                                         Network eaves. &
                                                                observe
                                                                DNS
                                                                or
                                                                proxy
                                                                traffic &
                                                               Customer Threat
    \\

        \S\ref{sec:authfail}: Payment bypassing / free use of pay service &
                                                  Internet user &
                                                                  send
                                                                  DNS
                                                                  resolution
                                                                  requests &
                                                           \sdns Provider Threat                                                                        
    \\

    \S\ref{sec:popularity}: Exposure to analytics / business analysis &
                                                Internet user &
                                                                  send
                                                                  DNS
                                                                  resolution
                                                                  requests &
                                                        \sdns Provider Threat
    \\

    \bottomrule
  \end{tabular}
\end{table*}
}

\Section{Introduction}

Vantage points matter on the Internet.  Websites often customize or restrict
content for clients based on their network locations and perceived geographic
locations. This is especially true of media streaming services such as Netflix,
Hulu, Pandora, and Amazon Prime Video, that are contractually obligated to
restrict audio/video content based on their users' geographic locations.  Such
websites establish so called {\em geofences} that enforce location-based access
control policies by geolocating clients based on their IP addresses.

However, determined users can apply simple methods to circumvent
geography-based
blocking by relaying connections through a proxy server
located within the fence. Commercial
VPN  providers describe such
abilities when marketing their services~\cite{ExpressVPN:online,NordVPN:online}. Free solutions such as
Tor~\cite{tor} and open proxies~\cite{openproxies,tsirantonakis2018large} also
enable users to bypass geofences. However, popular existing approaches
demand some user expertise and often require users to download and operate
specialized software.  
Worse, previous studies show that the use of open proxies may incur
severe security and privacy risks~\cite{openproxies,tsirantonakis2018large}.

There is a growing industry of {\em smart DNS} (\sdns) providers that
enable an interesting and unstudied method of circumventing
geofences. \sdns is simple and does not require additional software.  Instead, a
user reconfigures their computer's DNS settings to use an DNS resolver operated by a
\sdns service.  The \sdns resolver ``smartly'' identifies resolution requests for
restricted domains (hereinafter, {\em fenced} sites) and returns proxy
servers' IPs in lieu of these domains' correct IPs.  The client's machine then
directs its traffic to the specified proxy server (since that is the address to which the
domains resolve), which is located within the geofence.  Finally,
the proxy servers relay the clients' communication to and from these
requested domains.  For non-geofenced (hereinafter, {\em unfenced})
sites, DNS requests are resolved correctly.  Thus, the end-user needs only
browse as usual; all \sdns proxy management happens (potentially
unnoticed) without additional interaction.

This paper describes an exploration of \added{the privacy and security
properties} of smart DNS
services---to the best of our knowledge, the first such study in
the open literature. 
Through analyzing the architecture
and behavior of deployed \sdns systems, we provide  
descriptions of how \sdns services operate.

\added{Our analysis also uncovers several architectural weaknesses,
  implementation errors, and system misconfigurations that lead to pernicious
  privacy leaks, and are pervasive in the \sdns ecosystem:}

We 
 demonstrate a simple technique
by which any content provider could immediately identify both the use of
an \sdns service to access its site, as well as the actual IP address
of the requesting client.  
(We note that numerous content providers have already sought to crack
down on \sdns use, perhaps using the technique we describe here.)
This would allow the content provider to consistently identify 
the use of \sdns, without requiring them to continually discover and
block proxy servers, and, in so doing, engage in a never-ending ``whack-a-mole'' arms race with \sdns providers.

More troubling, we describe a design flaw
in the architecture of \sdns systems that enables content providers to
enumerate the IP addresses of an \sdns service's customers,
regardless of whether they are logged in to the service's web portal, or currently use one of its \sdns resolvers for their web browsing.
  And, as we show through proof-of-concept attacks, the implementations of some \sdns services allow any arbitrary third-party
to enumerate these \sdns services' customers.  We discuss in detail
the ethical considerations of our measurements and the proof-of-concept
attacks we conducted.

\added{We also identify a number of authentication and authorization errors,
coupled with misconfigurations, that effectively turn some \sdns
providers into a distributed network of open proxy servers.  That is,
we find that several \sdns providers fail to authenticate users who
access their proxies, and instead rely only on authentication at their
DNS resolvers.  We present simple methods for enumerating such open
proxies and explain how unscrupulous users could bypass paying for \sdns services
while reaping their benefits.}

We further find that some \sdns providers proxy more content than
advertised. \sdns providers do this by forwarding traffic for websites, for which they do not
advertise support, to proxy IPs. This raises the risk of content interception, manipulation, and eavesdropping, both by the \sdns provider and along the extended Internet path this traffic now traverses.

In addition to
exploring the privacy and security properties of \sdns services, we
also study the landscape of \sdns operators.
Our exploration of \sdns services, conducted over more than ten months, strongly suggests that the
\sdns marketplace may be more consolidated that it appears.  Several
of the identified \added{25} \sdns providers are actually the same entity
advertising their services under multiple distinct names and
websites. Our probes also exposed the popularity of different content
for \sdns providers as well as the \sdns providers themselves.
\iflongversion
Applying current virtual private server
(VPS) costs and advertised \sdns plan costs, we estimate
the costs and revenues of \sdns services, and find that they are
immensely profitable.
\fi

\Paragraph{Relevance to Privacy}
\changed{\sdns is provided by many existing VPN providers, perhaps due
  to overlap in infrastructure requirements, and \sdns is often
  advertised alongside VPN products.  The manner in which \sdns is
  marketed differs among providers, with some implying
  (falsely) that \sdns is itself a privacy-enhancing technology~\cite{ironsocket,vpnuk,vpnsecure}.
    We found no instances in which \sdns
  providers describe any added privacy risks.}

\sdns does not appear to be a niche industry.  At least two \sdns
providers (\htmlize{www.ibvpn.com} and
\htmlize{www.smartdnsproxy.com}) state that they have more than one
million users.  Our own measurements 
largely support this claim.

\changed{Our main findings---\sdns customer IP addresses can be easily
  mined by third parties; \sdns substantially increases users'
  vulnerability to eavesdropping; and content providers can trivially
  discover when users attempt to bypass their geofences---all threaten
  the privacy and/or security of \sdns customers.  Although \sdns may
  not itself be considered a privacy-preserving technology (although
 it is sometimes marketed as such), the
  architectural and implementation weaknesses we describe in this
  paper are relevant to the estimated millions of \sdns users, whose use
  of these systems may constitute significant and (until now)
  unexplored privacy risks.}

 \Section{Background on DNS}

\changed{DNS~\cite{dns:rfc} is the mechanism by which hostnames are
  mapped to IP addresses to facilitate Internet routing.  DNS is
  complex with several important nuances, but conceptually, DNS can be
  thought of as a distributed database, with mappings between
  hostnames and their IP addresses stored in {\em zone files}.
  Ordinarily, the
  owner of the domain (i.e., the party that registers the domain)
  effectively controls this mapping.}

\changed{Users {\em resolve}---that is, translate a hostname to its
  IP address---by querying a DNS resolver.  Typically, users use the
  resolver that is provided in the DHCP response they receive when
  joining a network; often, but not always, these resolvers are
  operated by the  ISP that provides Internet connectivity.  Users
  also have the option of selecting a different resolver: popular
  choices include Google's DNS and Cisco Umbrella's DNS resolver.}

\changed{When receiving a request, a resolver checks its cache for the
  queried hostname.  If it finds an unexpired entry, the cached
  results are immediately returned.  Otherwise, either the resolver
  returns a reference to another resolver ({\em an iterative query})
  or,  the resolver itself relays the request towards
  another resolver on behalf of the client ({\em a recursive query})
  and ultimately returns the resolved IP.  The resolver also caches a
  copy for a length of time that is defined in the corresponding zone
  file. The resolver that is responsible for a given domain is known
  as an {\em authoritative name server} and it is contacted in
  recursive queries when the answer is not cached by the other
  DNS resolvers.  We found that all \sdns resolvers support only recursive queries.}

\changed{While DNS supports both UDP and
  TCP, the former is much more common.   DNS is typically
  neither authenticated or encrypted. To address this and improve
  privacy and security, there are three main extensions to DNS that
  offer additional privacy features: DNSSEC, DNS-over-HTTPS
  (DoH)~\cite{doh:rfc} and DNS-over-TLS (DoT)~\cite{dot:rfc}. DNSSEC
  aims to ensure the authenticity of DNS data by incorporating a PKI
  and using signed and verifiable zone files. (Friedlander et al.~provide a good
  overview of DNSSEC~\cite{dnssec-explainer}.)  DNSSEC does not
  address confidentiality of the DNS request, merely authenticity.
  DoH and DoT, on the other hand, both provide confidentiality of DNS
  requests and responses by using TLS.  Importantly, \sdns is
  inherently incompatible with DNSSEC (since \sdns returns modified
  resolution results), and we found no \sdns providers that support
  either DoT or DoH.}
 \Section{Related Work}
\label{sec:rel}

There are large,
organized efforts at enumerating instances of Internet
censorship~\cite{encore,filasto2012ooni,echo-censorship} and there is
considerable work that examines methods of bypassing
blocking~\cite{tor,censorship-sok,censorship-empirical-study}.  However,
geo-filtering at the server-side is far less well-studied.

Afroz et al.~\cite{afroz2018exploring} performed a large-scale
measurement study and found that geo-filtering was ubiquitous on the
Internet.  A large number of commercial VPN providers, including many
of those listed in Table~\ref{tbl:costs}, advertise their services as
a means of getting around geo-fences.  Khan et
al.~\cite{vpn-ecosystem-imc2018} and Weinberg et
al.~\cite{proxies-geolocation-lying} independently analyzed the VPN
ecosystem, with both sets of authors concluding that VPN providers
regularly misrepresent the location of their endpoints.
Interestingly however, so long as the fenced website
similarly misattributes the VPN endpoint's location, this misclassification
 is not by itself problematic
for users wishing to defeat geofences.  Poese et
al.~measure the accuracy of geolocation services and find that errors
are fairly common~\cite{ip-geolocation-unreliable}.

Server-side filtering of clients has also been explored in the context
of preventing anonymous users (e.g., Tor users) from accessing
websites~\cite{zhang-ephemeral-exits,zhao-exitbridgeonions,khattak2016-torfiltering,singh2017-characterizingtor}.
The approaches used to detect access from anonymity
networks~\cite{khattak2016-torfiltering,singh2017-characterizingtor}
and countermeasures to bypass such
filtering~\cite{zhang-ephemeral-exits,zhao-exitbridgeonions} are
specific to the anonymity services being used and 
differ from the IP geolocation mechanisms used by content providers to
impose geofences.

Numerous efforts have attempted to map out and explore the performance
of the Internet's domain name system (cf.~studies by Callahan and
Allman~\cite{callahan2013modern} and Jung et al.~\cite{jung2002dns},
and measurement platforms such as the RIPE Atlas~\cite{ripe-atlas}).  We also measure
DNS performance
\iflongversion
(see Appendix~\ref{sec:performance}),
\fi
but focus in this paper on the added
costs incurred by choosing remote DNS resolvers.  Finally, DNSSEC
obviates the benefits of \sdns services by preventing the type of
forged DNS resolutions on which \sdns depends (we discuss the impacts
of DNSSEC in~\S\ref{sec:dnssec}).  However, DNSSEC has
seen slow adoption and even the resolvers that support DNSSEC often
fail to validate the authenticity of DNS
records~\cite{dnssec-ecosystem}, indicating that \sdns will likely
function for at least the short-term.

\added{As explained more fully in the next section, the proxies used by \sdns
providers inspect Server Name Indication (SNI) TLS
headers~\cite{sni-rfc} to extract the hostname requested by the
client.  Once the requested hostname is obtained, the proxies 
simply forward TCP traffic between the client and the destination.
Such proxies are often called {\em SNI proxies}, and have been
used as building blocks for domain fronting
systems~\cite{domainfronting} (e.g., Tor's
meek~\cite{tor,meek}) and more generally for proxying of Internet
traffic. Using ZMap~\cite{zmap} scans and a novel SNI proxy
testing tool, Fifield et al.~identify approximately 2500 {\em open} SNI
proxies~\cite{sniproxy-list} that service public requests.  We find that Fifield's list includes
some SNI proxies operated by \sdns services, highlighting these
services' failure to properly authenticate requests; we explore
authentication errors in more detail in \S\ref{sec:authfail}.
}

 \Section{Architecture of \sdns Services}  \label{sec:architecture}

There are two phases of \sdns usage: registration and operation.

During the {\bf registration} phase, a user must create an account on
the \sdns service via the service's webpage and, depending upon the
service, select a payment plan.  The user must also register her
public-facing IP address with the \sdns service.  Many services
simplify IP registration by presenting the detected IP address of the
user as the default (and sometimes only) option.
Next, the user must select a DNS resolver from a list of resolvers
operated by the \sdns service.  Many services advise the user to
select an \sdns resolver that is geographically located nearby.  This
reduces the network latency incurred during DNS lookups, which, as we
show in
\iflongversion
Appendix~\ref{sec:performance},
\fi
can significantly impact the user's
overall browsing experience.  Finally, the user must configure her
computer to use the selected \sdns resolver as its primary DNS
resolver. All \sdns services provide detailed instructions, complete
with screenshots, on how to carry out this process.

\begin{figure}[t]
  \centering
  \includegraphics[width=\linewidth]{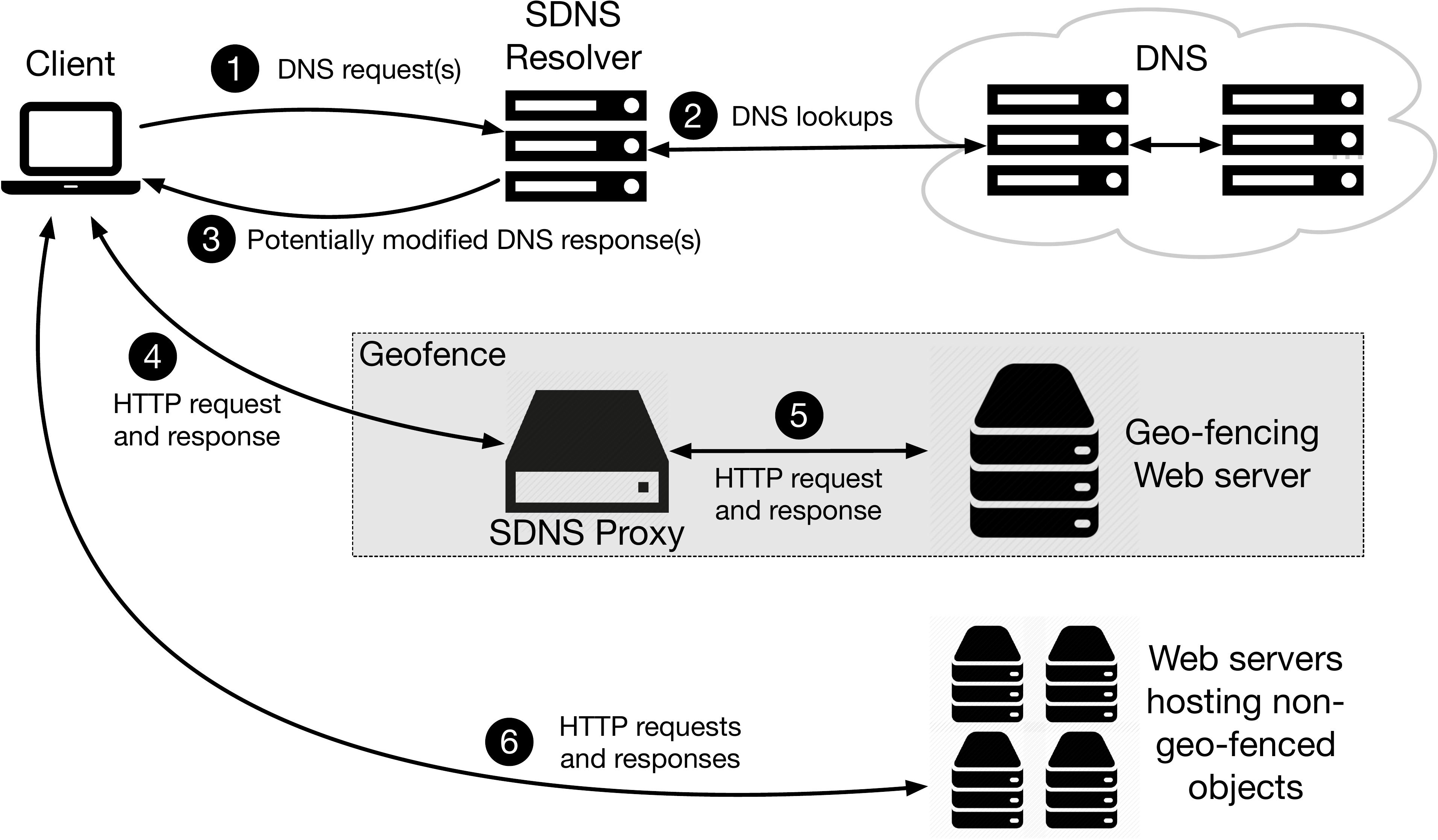}
  \caption{Workflow of the operation phase of a \sdns.}
  \label{fig:overview}
\end{figure}

The {\bf operation} phase is depicted in
Figure~\ref{fig:overview}. This begins when the user attempts to
access geofenced content that is supported by the \sdns service.  We adopt
the terminology of many of the \sdns services and refer to a
geofenced website proxied by an \sdns service as a {\em
  supported channel} or, more concisely,  as a {\em
  channel}.  (The term is likely inspired by TV channels; \sdns
effectively allows its users to ``tune'' to ``channels'' that would
otherwise be unavailable.)

Without loss of generality, consider a request for
\htmlize{https://netf\-lix.com}, a channel supported by the user's chosen
\sdns service.  The
user's DNS requests---for \htmlize{net\-flix.com} and for domains that host web objects
referenced on that page (e.g., \htmlize{fls.doubleclick.net})---are sent to the \sdns
resolver (step \circled{1}).  For each resolution request, the \sdns resolver
either returns the correct IP address (e.g., via recursive lookups, as depicted
in step \circled{2}) or returns the IP address of one of its proxies that
reside within the geofence (step \circled{3}).

It is worth highlighting that \sdns depends entirely on IP-based authentication
to determine whether the requesting user has completed the registration
phase. If the user is not registered, the \sdns resolver's behavior differs by \sdns provider. Most providers return a correct
(non-proxy) IP when resolving fenced content for non-customers.  (As we show in
\S\ref{sec:security}, doing otherwise can lead to serious privacy vulnerabilities.) 
\sdns cannot support more robust forms of authentication since (i)~DNS does not
support requestor authentication and (ii)~proxies cannot rely on web-based
authentication mechanisms, such as cookies; HTTPS prohibits the
proxy from inspecting session cookies, since TLS encrypts all content between
the client and the website.

Returning to our example of a registered \sdns user accessing
geofenced content, if the \sdns service supports \htmlize{ht\-tps://netflix.com}, the
user's configured device will send HTTP/S requests to the proxy IP  returned
by the \sdns resolver (step~\circled{4}).  The task of the proxy is twofold: first, it
must determine which site is being requested since a single proxy may serve
multiple channels (e.g., Netflix, Hulu, and ESPN).  If the request is over HTTP,
then the proxy can inspect the {\sc Host} HTTP header, which is mandatory in
HTTP/1.1.  For encrypted HTTPS traffic, \sdns exploits TLS' Server Name
Indication (SNI) extension~\cite{sni-rfc} that allows the requested site to be
communicated as cleartext.  SNI is intended to allow a single server to host
multiple domains and serve the correct TLS certificate during the
exchange.  SNI is a popular TLS extension and has been found
to be present in 99\% of TLS connections~\cite{frolov2019use}. 
 In the context of \sdns, the use of SNI allows the \sdns proxy to interlope on the
exchange and learn to which domain (e.g., Netflix) it should send proxy
traffic.

Second, the proxy must actually forward the traffic
(step~\circled{5}).  \sdns proxies operate transparently and function as
TCP endpoints for both the requesting client (where it poses as the
web server) and the web server (where it poses as the client).
\sdns proxies merely relay data received through one TCP
connection to the other, and vice versa; doing otherwise would disrupt
TLS (HTTPS) traffic between the client and the server.

The \sdns service does not necessarily have to proxy all web objects
that are included in a requested webpage (step~\circled{6}).  For example, the
SmartDNSProxy provider does not proxy
requests to \htmlize{fls.doubleclick.net}, even though such web objects are
referenced on \htmlize{netflix.com}.  This has the advantage that it decreases the proxy's
workload and operating cost.

\Subsection{DNSSEC and Encrypted SNI}
\label{sec:dnssec}
\sdns services are entirely incompatible with DNSSEC, since the latter
provides origin authentication of DNS records.  Of course, \sdns
resolvers do not support the DNSSEC extensions, making this
incompatibility moot until browsers and/or operating systems begin to
require DNSSEC support.

Cloudflare co-introduced and adopted~\cite{cloudflare-sni} encrypted
SNI~\cite{draft-ietf-tls-esni-03}, which eliminates a privacy weakness
of SNI by encrypting the requested hostname between the client and the
receiving web server.  Encrypted SNI would thwart the
current \sdns architecture by hindering a proxy's ability to identify
the site being requested. However, as of May 2020, encrypted SNI is either
unsupported or not enabled by default in the latest release versions
of Chrome, Firefox, Safari, Brave, and Microsoft Edge.

\Subsection{Contrasting with VPN Services}
A stark difference between \sdns services and VPNs is that the former
has no obvious on/off mechanism.  VPNs require starting an application
and authenticating to the VPN
provider.  In most settings, the VPN is not on by default, and there are
visual indicators on the desktop when the VPN is in use. User
intervention is also  required to reestablish the connection to the VPN
after machine reboots.  That is, the use of the VPN requires {\em intentional}
actions by the user.

In contrast, while \sdns providers offer their customers some helpful instructions and
tools to configure their DNS settings, \sdns is much less user-friendly with respect to activation or deactivation.
There are no obvious indicators (other than the availability of certain video
streaming services) that the \sdns service is in use.  Due to this
opacity in \sdns services, we posit that many \sdns users will  forget the
current status of their DNS settings and effectively {\em always} be performing
DNS lookups through the \sdns' resolvers.  We discuss the security and privacy
implications of continuously using \sdns services in
\S\ref{sec:security}, as well as the performance implications in
\iflongversion
Appendix~\ref{sec:performance}.
\fi

\sdns is also unlike VPNs in that \sdns does not encrypt content between the
user's device and the proxy.  It is unable to do so, since its use is entirely
invisible to the user's computer.  For non-HTTPS traffic, \sdns increases the
attack surface by allowing any potential eavesdropper between the
client and the proxy to perform man-in-the-middle manipulation.  In the case
of HTTPS traffic, an eavesdropper can inspect SNI headers to learn
which domain names are being requested.

\tabproviders{}

 \Subsection{\sdns Marketplace \pagebudget{.25}}
\label{sec:marketplace}

To understand the \sdns marketplace, we performed simple Google queries to identify potential providers.
\changed{We found that many \sdns providers are also VPN providers, advertising \sdns alongside VPN services and usually at a lower cost. \sdns, unlike VPNs, is not a privacy enhancing technology, but the commingling may confuse users about the security and privacy properties of \sdns.  In at least two instances (ibVPN and VPNUK), \sdns providers marketed \sdns alongside their VPNs as privacy-enhancing services.}

In total, we 
identified 25 \sdns providers. Using information available on their webpages, we catalogued (i)~their prices and subscription plan offerings, (ii)~the IP addresses of their DNS resolvers, and (iii)~the countries in which the providers appeared to be registered.

The names, monthly costs, and locations of the 25 identified \sdns providers are listed in Table~\ref{tab:providers}, \added{including 15 providers which we focused on as part of our analysis. (These providers were selected based primarily on their search rank when querying Google for \sdns providers and their costs.)}  The 25 \sdns providers spanned 12 countries, where the country of origin was determined by searching for contact information (i.e., mailing addresses) listed on the providers' web pages. When searching for listed contact addresses, we also noticed that a number of the Turkish \sdns providers mention on their respective webpages that they belong to  a single parent company.

\Paragraph{Company Aliasing}
\label{sec:aliasing}
\added{During our analysis of \sdns services, we gathered
  evidence that strongly suggests that some of the \sdns providers are
  in fact the same company advertising under multiple
  name brands.}

  We identify numerous instances in which \sdns
  providers share infrastructure.  To do so, we determine a
  (potentially incomplete) set of proxies used by each \sdns service
  by querying their DNS resolvers for supported channels (e.g.,
  Netflix), and then comparing the returned IP addresses with a large
  ground-truth dataset, which was obtained by resolving the hostnames from a
  distributed network of RIPE Atlas nodes~\cite{ripe-atlas}.  Our
  methodology for detecting shared proxies is described in more detail in
  Appendix~\ref{sec:measure}.

\added{We find that the SmartDNSProxy, Trickbyte, and Uflix \sdns services
  share extensive infrastructure; specifically, we identify 10 proxies that are used
  by all three providers, seven that are shared between SmartDNSProxy
  and Trickbyte, and 14 shared proxies between Uflix and
  SmartDNSProxy.  
  We additionally note that both Trickbyte and SmartDNSProxy's
   websites are served from the same /16 network,
  previously shared TLS certificate subjects, and are registered using the same
  domain registrar.}

\added{Upon additional inspection, we also discover evidence implying that CactusVPN and SmartyDNS are likely operated by a single
entity. Specifically, these two poviders share at least four proxies, and exhibit similar proxying behavior patterns, which we describe in more detail in \S\ref{sec:authfail}. }

\added{The rationale for operating multiple seemingly (but not
  actually) distinct \sdns services is unclear.  We conjecture that
  such a strategy may attract more customers, since there are several
  sites that feature reviews and rankings of \sdns
  services.\footnote{See, for example
    \url{https://thevpn.guru/top-smart-dns-proxy-providers/} and
    \url{http://www.bestsmartdns.net/}.}
  Operating as multiple services increases the chances of appearing at
  the top of at least some rankings.  This is similar to findings that
  multiple VPN services may be operated by the same
  entity~\cite{vpn-ecosystem-imc2018}, perhaps also to gain
  advantage in VPN review and ranking sites.}

 \Section{System and Threat Models}
\label{sec:models}

\changed{There are several actors in the \sdns ecosystem: {\em \sdns
    providers} sell geofence-evading services to {\em customers} in
    order to provide more unfettered access to geofenced {\em content
    providers}.  (We also refer to customers as {\em users}.)  The
    \sdns infrastructure is composed of one or more provider-operated
    {\em resolvers} and one or more {\em proxies}.  Additionally, the
    \sdns resolvers depend on the {\em traditional DNS
    infrastructure}, since customers' DNS queries correspond not just
    to supported content providers (e.g., netflix.com), but also to
    unproxied domains (e.g., petsymposium.org).}

\tablethreats

\changed{This paper explores the privacy and security implications of
  \sdns to both customers and
  \sdns providers, and thus we consider two separate threat models:}

\Paragraph{Customer Threats} This paper considers attacks on
  customer privacy that expose a customer's IP address, either to the
  content provider or to any outside party.  We note that such
  exposure could potentially present legal risks to \sdns
  users\footnote{In the U.S., the use of \sdns may technically violate the
    Computer Fraud and Abuse Act, which criminalizes
    ``exceed[ing] authorized access'' of a computer system and imposes
    civil liability on the perpetrator~\cite{cfaa}.}, or result in
  users being banned by content providers.

  It is worth emphasizing that, as with other work
  (cf.~\cite{tor,i2p}), we consider IP addresses to be sensitive
  information.  Indeed, the EU's General Data Protection Regulation
  (GDPR) and the California Privacy Protection Act of 2018 both
  consider a user's IP address to be personally identifiable
  information under certain circumstances~\cite{gdpr,ccpa}.  We
  describe the state-of-the-art in mapping IP addresses to specific
  individuals in Appendix~\ref{sec:ip2person}, but highlight here that
  IP-to-individual mappings are commercially available (e.g., from
  Experian) and that IP-to-individual search engines are also
  available (e.g., \url{https://thatsthem.com/reverse-ip-lookup}).

  Additionally, we argue that the exposure of customer IP addresses to
  {\em any} arbitrary outside party falls well outside of the norms
  that users expect from their Internet services.  We can think of no
  other example in which a service allows outside parties to enumerate
  all of its users.  Perhaps more importantly, we did not find any
  \sdns service that advises its customers about any such exposure.

  Finally, we consider customers' increased vulnerability to traffic
  analysis due to the use of \sdns.  We consider the
  additional risk not just to traffic directed towards content
  providers (which would take longer paths due to proxying) but also
  other more general Internet traffic that users may not expect to be
  proxied.

\Paragraph{Provider Threats} We also explore the privacy
  and security risks of operating an \sdns service.  
  These consists of vulnerabilities that either (1)~harm the operation
  of the \sdns provider or (2)~reveal potentially sensitive
  information about its operation.

  More concretely, such threats include fundamental weaknesses in the
  \sdns architecture that allow a content provider to detect (and thus
  block), in real-time, the use of \sdns.  (We note that this is a
  more powerful attack than attempts to enumerate \sdns proxies, since
  it avoids an arms race between discovering proxies and spinning up
  new proxies.)  Additionally, we consider to be in-scope attacks that
  target the financial operation of an \sdns service and allow users
  to bypass payment and effectively access the service for free.

  Finally, our threat model includes the exposure to analytics that
  enables outsiders to perform competitive analysis on the \sdns
  provider.  This includes the ability of a third-party to estimate
  the number of users and revenue of an \sdns service, as well as to
  gauge the relative popularity of the channels that it proxies.

\Paragraph{Adversaries and adversarial capabilities}
We consider several adversaries that pose threats to either customers
or providers.  Our {\bf content provider adversary} operates a channel
that is targeted for geofencing bypassing by an \sdns provider.  The
content provider has the ability to modify its website and inspect its
own web logs.

The {\bf network eavesdropper adversary} is a passive network
observer.  We consider two variants of our network eavesdropper: an
eavesdropper who is located between the client and the client's \sdns
resolver, and a network eavesdropper that is located between the \sdns
proxy and the destination (geofenced) website
(see~\S\ref{sec:eavesdropping}).  An example of the former is the
\sdns user's ISP; an example of the latter is a government or AS that
monitors or hosts the proxy server.  The network eavesdropper can
inspect intercepted packets.  For the eavesdropper who observes DNS
traffic, our attack is effective when the DNS request and response are
not encrypted.  We are not aware of any \sdns service that supports
encrypted DNS resolution (i.e., with either DoT~\cite{dot:rfc} or
DoH~\cite{doh:rfc}).

We also present a number of attacks that can be carried out by nearly
any arbitrary third-party Internet user; we call this adversary the
{\bf Internet user adversary}.  The Internet user adversary can
exploit the authentication and authorization failures we identify in
\S\ref{sec:authfail} to use \sdns services without having to pay for
them.  We show that such attacks are possible and can be carried out
by any Internet user.

An Internet user adversary can also probe the caches of \sdns
providers' DNS resolvers to infer information about the popularity of
the \sdns providers as well as which channels are most often used by
the providers' customers (see \S\ref{sec:popularity}).  Here, we
require that our Internet user adversary be able to identify the DNS
resolvers used by an \sdns provider.  We note that resolvers are
listed on \sdns providers' websites since \sdns customers need such
information to configure their computers to use \sdns.

Finally, the Internet user adversary can carry out the customer
enumeration attack (see \S\ref{sec:enumeration}).  Here, three
additional capabilities are needed: the adversary needs to be able to
(1)~register a domain name (of the adversary's choosing), (2)~operate
the authoritative name server for that domain, and (3)~be capable of
sending spoofed UDP packets.

\medskip      

\noindent We summarize our main security and privacy findings in
Table~\ref{tbl:threats}.  We note that our threat models exclude
geofence circumvention. Although this may be reasonably considered a
security threat to content providers (since it bypasses an
authentication check), this is the intended function of \sdns
services. Our focus in this paper, rather, is to shed a light on the
previously undocumented privacy and security risks of \sdns.

 \Section{Privacy Vulnerabilities in \sdns Designs and Implementations}
\label{sec:security}

\sdns' architecture and implementations lead to several privacy and
security risks, which we describe below.  \changed{For each
  vulnerability, we list the relevant threat model
defined in~\S\ref{sec:models}.}

\Subsection{Client Enumeration Attacks}
\label{sec:enumeration}
\threatmodel{Customer}

Standard DNS does not support client authentication, and hence \sdns
providers must rely on IP-based authentication to identify customers.
The use of IP-based
authentication, coupled with the ease at which UDP-based DNS requests
can be forged leads to serious
privacy vulnerabilities. (All tested \sdns services support UDP-based DNS.)

We discovered architectural weaknesses in two \sdns services that
allow a third-party attacker (the {\em Internet user adversary}
described in \S\ref{sec:models}) to query the \sdns service and, in so
doing, determine whether a target IP address belongs to one of its
registered customers. When repeated, this attack allows the attacker
to enumerate the IP addresses of these services' customers. For ease
of exposition, we refer to an IP address associated with a customer of
the \sdns provider as being a {\em registered} IP.  The attack
requires no client interaction and will reliably reveal whether an IP
address is registered even if the customer is not actively using the
\sdns service, or even if it is not currently online.

As discussed in \S\ref{sec:models}, an adversary who learns the IP
addresses of \sdns users could potentially also combine this information
with existing IP-to-individual to determine the users' identities.
This, in turn, could enable targeted cease and desist notifications.  
Even without resolving particular identities, knowledge of \sdns
users' IP addresses is sufficient to deliver abuse notifications
to the operators of the users' networks (e.g., their ISPs), akin to
how movie and music trade associations communicate their perceived
violations of the U.S.~DMCA.

The client enumeration attack requires only that the adversary (i)~registers a domain
name (of the adversary's choosing) and (ii)~operates its own
authoritative domain server for that domain.  The adversary can be
located far from the \sdns service and need not intercept any messages
destined to \sdns resolvers.  While the attacker can be any Internet
user who meets the above criteria, we posit that content providers,
trade associations, and content producers (or their copyright holders)
might be especially motivated to enumerate the users of \sdns.

The attacker exploits a specific \sdns behavior in which the service's
DNS resolvers send distinct responses to customers' and non-customers' requests.
At a high level, the adversary uses these two different behaviors
to deduce whether an arbitrary IP address is a customer of the service or not.
This process can then be repeated for all IPv4 addresses (or more likely, a target set of IP addresses for which the attacker is interested).

\begin{figure*}[t]
  \centering
  \begin{minipage}[t]{.45\linewidth}
    \centering
    \includegraphics[width=\linewidth]{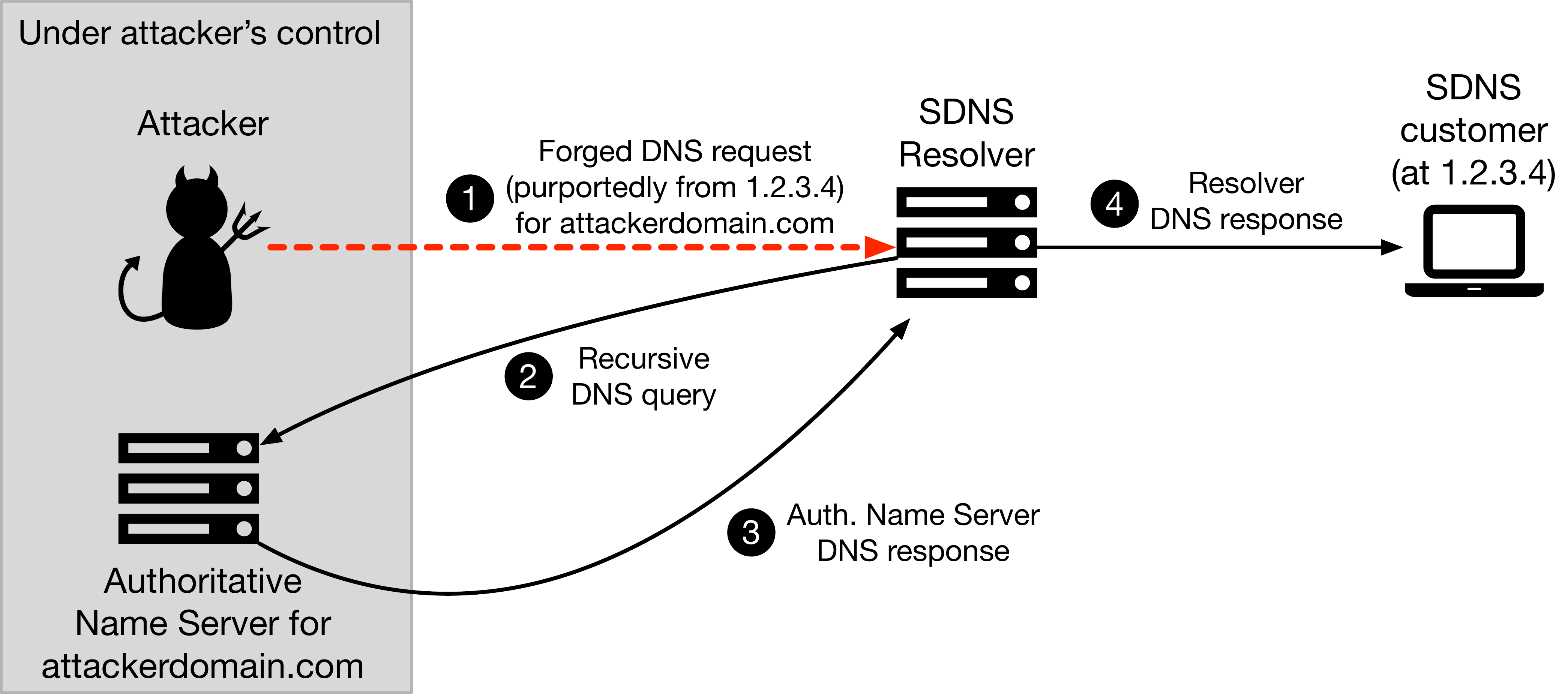}
  \end{minipage}
  \hfill
  \begin{minipage}[t]{.45\linewidth}
    \centering
    \includegraphics[width=\linewidth]{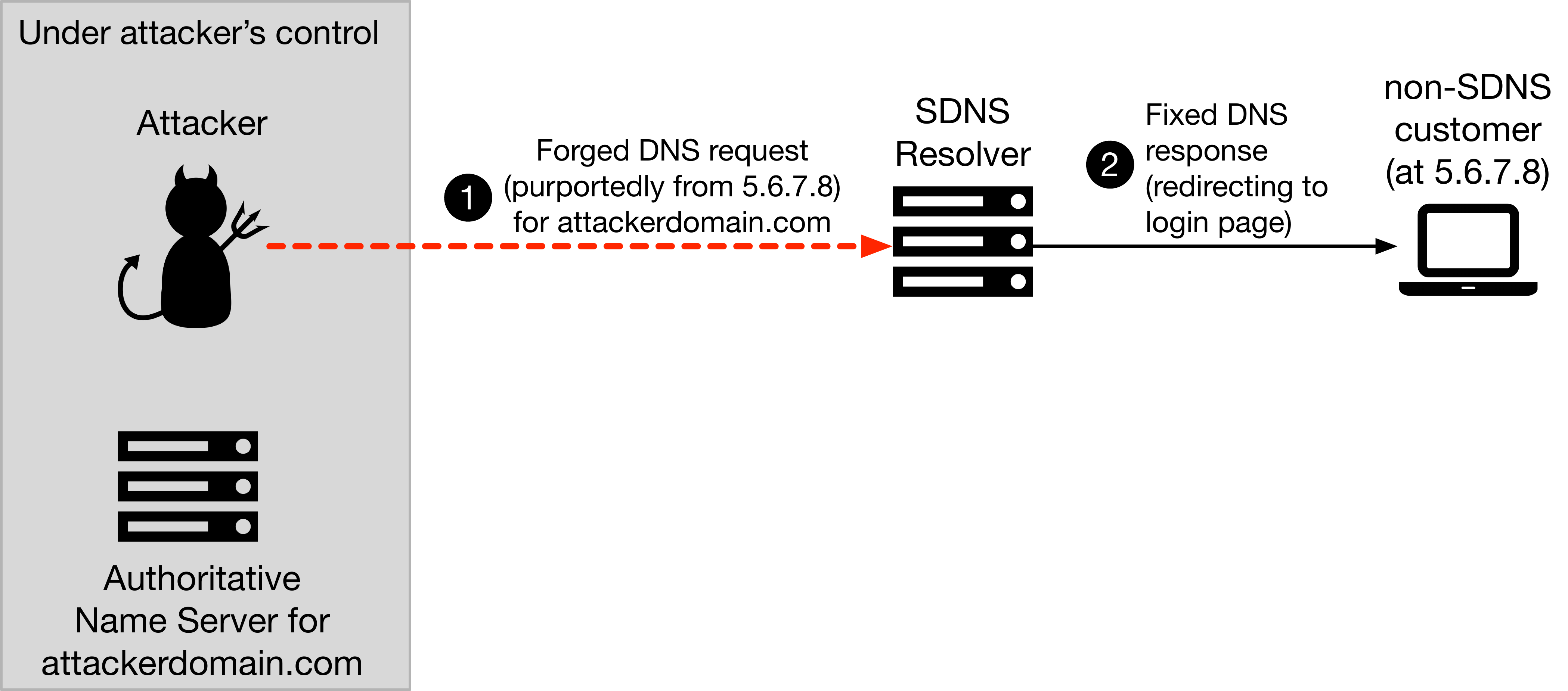}
  \end{minipage}
  \caption{The two possibilities for the client enumeration attack:
    either the candidate IP address belongs to an \sdns customer {\it
      (left)} or not {\it (right)}.}
  \label{fig:enumerationattack}
\end{figure*}

Figure~\ref{fig:enumerationattack} presents an overview of our attack.
To determine whether an arbitrary IP address, say \htmlize{1.2.3.4},
is \protect\linebreak registered, an attacker who controls
the domain \htmlize{attackerdomain.com} forges an otherwise well-formed DNS
request to the \sdns resolver for \htmlize{nonce.attackerdomain.com}, purportedly originating from \htmlize{1.2.3.4} (step \circled{1} in
Figure~\ref{fig:enumerationattack}, {\it left}), where \htmlize{nonce}
is a unique identifier. If \htmlize{1.2.3.4} is a registered IP address, the forged query for \htmlize{nonce.attackerdomain.com} would cause the \sdns resolver to 
correctly resolve \htmlize{nonce.attackerdomain.com} to its IP address (step \circled{2},
{\it left}) via recursive DNS lookups.
This would be the case, as the hostname \htmlize{nonce.attackerdomain.com} does not correspond
to any channel supported by the service.

We emphasize that to support general web browsing, \sdns
resolvers must correctly resolve hostnames for domain names they do
not support.  Additionally, the use of a unique nonce prevents
\htmlize{nonce.attackerdomain.com} from being cached at the resolver.  This ensures that
the request is propagated to the authoritative name server for
\htmlize{attackerdomain.com}, where it can be observed by the adversary.  Finally,
the IP address for \htmlize{nonce.attackerdomain.com} (or an error if not found) is relayed back to the \sdns provider's resolver
(step \circled{3}, {\it left}) and forwarded onto the forged IP
address $X$ (step \circled{4}, {\it left}), where it is likely
discarded.

The right-side of Figure~\ref{fig:enumerationattack} shows the
alternative case in which the IP address 
\htmlize{5.6.7.8} is tested and is not registered with the \sdns provider.
 Here, we rely on a particular behavior of certain \sdns
providers; namely, that they  {\em do not} resolve requests from non-customers.  We found
two slight variations of susceptible behavior.  The ibVPN \sdns service
responds to non-customer hostname resolution requests by returning 
a fixed IP address belonging to a website it operates;
the website redirects the user to an error page.
This scenario is depicted in step~\circled{2} in Figure~\ref{fig:enumerationattack}, {\it right}.  In
contrast, the VPNUK service does not respond at all to non-customer
DNS resolution requests.

Both behaviors allow the attacker to determine whether an arbitrary IP
address, in this case \htmlize{5.6.7.8}, is a customer: if it is, it will receive the recursive
lookup request from the \sdns resolver and can observe this request at
its authoritative name server; if \htmlize{5.6.7.8} is not a customer, the request
will not appear.

To validate our attack, we performed a proof-of-concept experiment
using the ibVPN and VPNUK \sdns providers.  We discuss the ethics of
our experiment in \S\ref{sec:ethics}.  Our procedure was
identical for both systems: we purchased an account on the system and
registered our client IP address.  We also purchased a domain
name and configured an authoritative name server (hosted
at Georgetown University) for that domain.  We confirmed that requests
originating from our client's IP to resolve a unique subdomain were
recursively resolved and observed at our authoritative name
server.

Next, we confirmed that requests sent from an unregistered IP
(also operated by us), either yielded false static responses
(ibVPN) or no responses at all (VPNUK); in both cases, the requests
originating from the other, non-registered IP address did not propagate back to our authoritative
name server.

Finally, to complete the attack, we acted as the attacker using a
third IP on a different network.  The attacker forged two requests:
one purportedly from the registered IP and one from the non-registered
IP.  
We confirmed that only the forged requests that purported to be from the registered IP address were relayed to our authoritative name server.

\Paragraph{IPv4-space enumeration} We used the ZDNS tool from the
ZMap Project~\cite{zmap} to estimate the service capacity of our
institution's (i.e., Georgetown University's) DNS server.  ZDNS performs highly parallel DNS lookups
using lightweight Go threads, and is useful for efficiently resolving
a large number of domains against a DNS resolver.  We emphasize that
we did not use ZDNS against the ibVPN or VPNUK resolvers since ZDNS
could potentially disrupt their services.  We use the performance
measurements of our institution's DNS resolver only to form a rough
estimate of the length of time it would require to enumerate all
$2^{32}$ potential IPv4 addresses.

We find that Georgetown's DNS resolvers can resolve the top
10,000 Alexa sites in 7.462s (1340.12 requests/second), while
consuming less than 1~MBps.  At this rate, it would require
approximately 5.3 weeks of continuous queries to enumerate all
possible customer IPs (again, under the assumption that the \sdns
provider's resolver has comparable performance).  While such a
sustained rate of access is likely unrealistic, we note that large ISPs can be fully enumerated
in under a day (e.g., Comcast has approximately 71 million IP
addresses~\cite{comcast:ipinfo}). 

\Paragraph{Mitigations} Our attack relies on \sdns resolvers that
resolve an attacker-controlled domain only when the (purported)
requester is a customer.  The attack can be partially mitigated by
consistently and correctly resolving domains for all hostnames that
are not associated with a supported channel.  Indeed, one day after we
disclosed our attack to ibVPN, the ibVPN service implemented this
mitigation.

We emphasize that although this fix disallows arbitrary third-parties
from enumerating customers, it will not prevent the operators of a
supported channel (e.g., Netflix) from carrying out the attack.  For
example, the content operator can register subdomains (e.g., {\it
  nonce}.netflix.com) and forge a DNS request from a candidate IP $X$
to determine if $X$ is associated with a customer of the \sdns
service.  The \sdns provider cannot apply the above fix here, since to
route around geofences, it needs to respond to the client with an incorrect
resolution (containing the IP address of a proxy) when the requested
site is a supported channel.  

A more robust mitigation is for the \sdns resolver to accurately
resolve all resolution requests.  When the requested hostname
corresponds to a supported channel, the \sdns resolver can ignore the
correct IP address and instead return the address of its proxy to the
requesting client.  However, while this fully mitigates the attack, it
also allows a content provider (i.e., channel operator) with knowledge
of the \sdns service to precisely measure how often the \sdns service
is being applied to bypass its geofilter: it can inspect its own
authoritative name server's logs for relayed requests from the \sdns
resolver.

\Subsection{De-proxying by the Content Provider}
\label{sec:deproxying}

\threatmodel{Customer \& \sdns Provider}

It is relatively straightforward for a geofenced content provider
(i.e., a website operator) to (i)~detect and prevent access from an
\sdns service and (ii)~identify the true IP address of the \sdns
customer. A {\em de-proxying attack} requires the content provider to
insert content into its web page that {\em does not require a DNS
  lookup.} By causing DNS resolution to be skipped, the content
provider prevents the use of a proxy and forces the client to perform
a direct access.

Without loss of generality, consider a content provider
\htmlize{istream\-videos.net}, where \htmlize{istream\-videos.net}
resolves to the IP address \htmlize{1.2.3.4}.  To perform a
de-proxying attack, the content provider serves the partial content
\htmlize{\textless IMG
  src="https://1.2.3.4/ima\-ge.\-jpg?\-session\_id"\textgreater} where
\htmlize{session\_id }is a unique tag that can link the web requests
to \htmlize{istreamvideos.net} with those to
\htmlize{1.2.3.4}.\footnote{Although it is not particularly common (or
  advised), some certificate authorities (e.g.,
  GlobalSign~\cite{globalsign}) will issue IP-based certificates.}
The client's browser will process the above IMG tag and {\em directly}
access the image at \htmlize{1.2.3.4} since the (unused) \sdns
resolver loses the opportunity to return the IP of a proxy.  The
content provider then checks whether the two linked requests
originated from the same IP; significantly differing requesting IP
addresses (e.g., from different autonomous systems) indicates the use
of an SDNS service.

As a proof-of-concept, we performed a de-proxying attack against ourselves, using the
Hide-My-IP \sdns service.  Hide-My-IP proxies {\em all} connections:
its DNS resolver returns the IP address of a single proxy regardless
of the requested domain.  When the proxy receives HTTP requests from
the client, it inspects the {\sc Host} HTTP header or the SNI TLS
header to identify the requested destination, and then acts as a
transparent TCP proxy.  Since Hide-My-IP proxies all sites, we can
trivially become a ``content provider'' by simply instantiating a web
server.  As described above, we constructed a simple web page that
included an IMG tag whose source (``src'') was specified by IP rather
than hostname.  We confirmed that our web server logs revealed that
the domain-based request for the webpage had a different requesting
client than the one for the IP-specified image; the former showed the
proxy IP address and the latter revealed the client's IP address.

\changed{The deproxying attack enables a content provider to learn
  which of its users use \sdns.  Unlike the client enumeration attacks
  described in~\S\ref{sec:enumeration}, the deproxying attack may
  directly implicate a user of the content provider if the provider
  requires users to first log in before accessing content.  It also
  provides a real-time mechanism for {\em immediately} detecting the
  use of \sdns, and is thus a far more practical means of protecting
  against geofence circumvention than frequently enumerating all \sdns
  users. Once an \sdns user has been identified, the content provider
  could either suspend or terminate that user's account, or simply
  disallow use of the service while \sdns is in use.}

There are no clear mitigations to a de-proxying attack, and moreover,
de-proxying attacks are particularly worrisome for users who
misunderstand the privacy properties of \sdns services.  While \sdns
services do not advertise anonymity, end-users could be confused about
the kinds of protections (or lack thereof) that these services
provide, especially when these same providers sell VPN services as
their primary offering.  This confusion could put end-users in
restrictive regimes at particular risk, if they access censored
content with an expectation that their accesses are anonymous.

\changed{The deproxying attack also presents a threat to \sdns
  services.  We found instances in which content providers blocked
  access from a handful (but not all) \sdns proxies.  This indicates a
  ``whack-a-mole'' defense in which content providers attempt to
  identify and block proxies.  This arms race generally works in the
  \sdns provider's favor, since cloud-hosted proxies can easily change
  IP addresses.  (This same whack-a-mole strategy is also used to find
  VPN services' egress points.)  The deproxying attack avoids this
  arms race by identifying in real-time the use of \sdns, and thus
  enabling immediate discovery of \sdns proxy servers as soon as they
  are utilized.  In short, the content provider can apply this attack
  to entirely eliminate the utility gained by using an \sdns service.}

\Section{Susceptibility to Eavesdropping}
\label{sec:eavesdropping}

\threatmodel{Customer}

\sdns services increase their users' susceptibility to eavesdropping.
We explore this increased risk across two dimensions: eavesdropping on
DNS requests and eavesdropping on proxies.

\Subsection{Eavesdropping on DNS requests}
A log of DNS queries provides a fairly complete record of which sites
and services were accessed by a requestor.  \sdns customers configure
their computers to send {\em all} DNS queries to \sdns resolvers, regardless of whether the queries pertain to
fenced or unfenced websites. This provides \sdns services with a 
comprehensive set of potentially sensitive metadata about their
customers.  We emphasize that this is in stark contrast to using VPNs,
whose use can be easily toggled on and off and whose active use is
typically indicated by visible cues presented to the user.  That is,
the ``set and forget'' configurability of \sdns services, described in \S\ref{sec:architecture}, has important
implications to users' privacy.

\Paragraph{Longer paths increase susceptibility to eavesdropping}
The architecture of \sdns services risks exposing
their users' Internet metadata to third parties, beyond the SDNS provider.
DNS requests and responses are usually (and, in the case of \sdns, we
believe always) sent unencrypted\footnote{The Firefox web
  browser now uses encrypted DoT~\cite{dot:rfc} to Cloudflare's DNS resolvers by default. However,
  \sdns users would need to disable this setting.}, allowing
eavesdroppers  between the client and the resolver
to learn which hostnames are being requested, and by
whom.

For regular (non-\sdns) DNS resolution, the resolver is typically
located near the requestor, and is often operated by the requestor's
ISP, which we note, learns the sites being requested by virtue of
forwarding their traffic.  That is, using a local resolver poses
little additional privacy risk.  Public DNS services, such as those
offered by Google, Cloudflare, and Cisco, serve as popular alternatives to relying on local
resolution, especially among more technically sophisticated users.  We
emphasize that the public DNS infrastructure offered by Google,
Cloudflare, and Cisco all use IP anycast and are backed by highly
distributed networks~\cite{anycast8888}.  For example, DNS resolution
requests to the fixed IP address of Google's Public DNS resolvers
will often be routed to a resolver located close to the
requesting client~\cite{greschbach2017effect}.

\tabletraceroute

Compared with local DNS resolution and with resolution via large,
public DNS providers, resolution via \sdns \changed{resolvers causes
  the} requests (and responses) to transit longer network paths.
\changed{We confirm this by counting the number of autonomous systems
  (ASes) traversed between clients and (1)~Google's and Cloudflare's
  public resolvers and (2)~108 identified \sdns resolvers.  We
  determine the number of ASes by performing traceroutes and using the
  utility's built-in IP-to-ASN mapping, and then counting the unique
  ASes observed in the reported network paths.  More AS traversals
  indicate longer paths and consequently increased vulnerability to
  eavesdropping, since more organizations have the ability to observe
  the traffic.  We place our traceroute clients in five 
  continents, and report our results in
  Table~\ref{tbl:traceroute}.  We find that the average number of AS
  traversals between the client and its chosen DNS resolver increases
  when the client elects to use a \sdns resolver.  The relative
  increase in the number of AS traversals ranges from 13\% (Japan) to
  a near tripling in length (Belgium), relative to using Google's or
  Cloudflare's public DNS service; in the United States, the average
  number of ASes that observe the DNS requests increases by 55\% when
  \sdns is used.}  \changed{Finally, we note that the} use of distant
DNS resolvers has been found to be a significant threat to privacy in
the context of Tor~\cite{greschbach2017effect}.  \changed{We
  emphasize, however}, that unlike with Tor, \sdns users send {\em
  all} DNS requests to potentially distant DNS resolvers, not just
those that are produced when temporarily browsing anonymously with a
specialized browser.

\Subsection{Eavesdropping on Proxies}
\label{sec:overproxying}
\sdns customers are also exposed to an increased risk of eavesdropping through the use of the proxies themselves. 
Communicating via a proxy increases the surface area for eavesdropping.  While directly accessing sites  generally uses the shortest paths in terms of the number of autonomous systems traversed~\cite{idmaps}, relaying traffic through an \sdns proxy requires that it first be transmitted to the  proxy and that the proxy separately transmit it to its destination. This process produces longer paths that are more vulnerable to eavesdropping.  

These long paths are especially risky in the case of \sdns services since connections between users and their proxies are not encrypted.  A user may use HTTPS to achieve end-to-end confidentiality of content  with the visited website, but the widespread use of SNI allows an eavesdropper situated either between the user and the proxy or between the proxy and the destination to learn \changed{the  hostnames of all requested URLs}.

  \changed{How much the eavesdropper can learn from this leakage mainly depends on what traffic the \sdns provider proxies. 
  	At the extreme, the HideMyIP \sdns service proxies all web requests, regardless of the requested webpage. 
  	Clearly, this leaks significant information to an on-path, passive eavesdropper and causes the \sdns provider to incur a very high bandwidth cost.}  

      \changed{However, even in cases where \sdns providers take steps to limit unneccessary proxying, they likely still leak substantial information about their users' Internet browsing habits.  The \sdns provider ultimately decides which domain names it will route to its proxies and which it will allow its users to access directly. As noted in~\S\ref{sec:architecture}, \sdns services can limit unnecessary proxying by only proxying the content required to make their  supported channels work---for example, just those web objects that consider the client's location and enforce the geofence.  
This can become problematic when a supported channel runs its geo-ip checks on a large CDN and references it using a ubiquitous domain name.  In one such case, we noted that Netflix runs one of its geo-ip checks on an Akamai node (akamaihd.net).  This effectively requires the \sdns provider to proxy {\em all} content to akamaihd.net (including that which is not related to any supported channel).  For example, the SmartDNSProxy \sdns service proxies some Akamai-hosted webobjects on the Honda motorcars website, despite Honda not being a supported channel.}

      \changed{This ``over-proxying'' allows an eavesdropper situated between the client and the proxy to learn not only about visits to supported channels, but also other sites that happen to use the same CDN nodes as those channels.  We note that although such information may be encrypted, the now-ubiquitous use of SNI may leak information about requested hostnames.}

\Paragraph{Unadvertised proxying}
\changed{Given \sdns providers' opportunity to limit costs by only proxying domain names needed to support their advertised channels, we expected \sdns providers to support only the channels that they advertise.
However, we additionally identified several instances in which this was not the case.}
  To discern instances of unadvertised proxying, we queried \sdns resolvers for domains from the Alexa website rankings list, and then compared the results to a ground-truth dataset we collected by issuing queries from a distributed collection of RIPE Atlas proxies as well as queries using Google's, Cloudflare's, and our local institution's DNS resolvers. (We exclude HideMyIP from this analysis, since it proxies all connections regardless of destination.)  A more detailed explanation of our methodology is presented in Appendix~\ref{sec:measure}.

We find that
four  \sdns providers (SmartDNSProxy, TrickByte, Uflix and VPNUK) omit supported domains 
from their published channel lists.

For the most part, the uncovered unadvertised channels correspond to
pornographic websites. We posit that these sites are
intentionally omitted from \sdns providers' websites to avoid 
detracting from the providers' perceived legitimacy or professionalism.
As with over-proxying of domains, the failure of \sdns providers to announce the proxying of
these channels poses privacy risks to their customers, since accessing these sites likely traverses longer Internet paths than would occur via direct connections.
\changed{Beyond the longer paths incurred, all proxied sites leak the SNI hostname of websites visited over a TLS connection. As such, the aforementioned (passive, on-path) eavesdropper can learn that these users access pornographic sites, which sites they visit, and the frequency at which they do so. Moreover, the \sdns provider can change which domains it proxies at any time, and without warning. As a result, the eavesdropper could potentially gain insights into additional aspects of users' browsing behavior.}

 \Section{Authentication and Authorization Failures}
\label{sec:authfail}
\threatmodel{\sdns Provider}

As discussed in \S\ref{sec:architecture}, the workflow of \sdns is a two-step process: (i)~upon receiving a DNS resolution request for a supported channel, an \sdns resolver returns the IP address of one of its proxies, and (ii)~upon receiving HTTP/S requests from the end user, the proxy then either inspects the \htmlize{Host} HTTP header or the TLS Server Name Indication (SNI) extension field to infer the destination, and then forwards TCP traffic to and from the location inferred.  Critically, \sdns providers should perform authentication  ({\em is the requesting user a registered customer?}) and authorization ({\em should the traffic to the requested site be proxied?}) at both of the above steps.

In prior work, Fifield et al.~used custom ZMap scans~\cite{zmap} to identify approximately 2500 {\em open SNI proxies} in the wild that used SNI introspection to proxy HTTPS traffic for arbitrary Internet users~\cite{sniproxy-list}.  We compared our list of identified \sdns proxies to the open SNI proxies found by  Fifield et al., and discovered four IP addresses on both lists, indicating that at least some proxies operated by \sdns services do not properly perform authentication.  That is, they allow non-paying customers to directly use their proxies to relay traffic (e.g., to bypass geofences).  We confirmed that the \sdns proxies allowed non-registered Internet users to proxy content by manually setting the SNI header in HTTPS requests originating from an IP that is not registered with the \sdns service. In all cases, the proxies retrieved the requested content.

To explore whether authentication failures were due to infrequent configuration errors on a small subset of a service's proxies or endemic misconfigurations across all of its proxies, we identified additional proxy servers for eight \sdns providers (see Table~\ref{tab:ouproxies}).  To find additional proxies, we noted that \sdns proxy servers sometimes presented distinctive error messages when accessed directly over HTTP, without a modified \htmlize{Host} HTTP or SNI header that indicated an alternate destination.  (Fifield et al.~made a similar observation of open SNI proxies~\cite{sniproxy-list}.)  These error messages often warned the user of a DNS misconfiguration and directed her back to the \sdns provider's website.  We queried censys.io for this distinctive text to discover more potential proxies.

Table~\ref{tab:ouproxies} shows the number of proxies we identified for each service, along with whether the proxies were restricted to \sdns customers (open circles) or functioned as open proxies (darkened circles).  We tested whether a proxy was open by specifying \htmlize{Host} HTTP headers (for non-encrypted web traffic) and TLS SNI headers (for encrypted web traffic) in an attempt to proxy.  We found that CactusVPN, HideIPVPN, and SmartyDNS all had endemic authentication errors; all of their proxies functioned as open SNI proxies for any requesting Internet user.  Oddly, we found no instances of open proxying for unencrypted traffic, even among those three providers.  The authentication checks---which are based solely on the requestor's IP address---are performed only for HTTP proxying.  We note that Google reports that between 74 to 94\% of web requests using the Chrome browser (depending upon computer platform) are over HTTPS~\cite{transparencyreport}, suggesting that the failure to authenticate HTTPS requests is sufficient for non-customers to access the majority of the web.

We additionally checked whether the identified proxies would forward traffic to any domain, or limit proxying to its supported channels (as determined by its responses to DNS resolution requests with a proxy's  IP address).
We use the term {\em universal} to refer to proxies that forward traffic to any domain, and note that their presence in a given \sdns provider's infrastructure indicates the provider's failure to properly check the authorization of its proxying requests.

\tableproxyuserauth{}

As shown in Table~\ref{tab:ouproxies}, we find that the identified proxies operated by CactusVPN, HideIPVPN, ibVPN, SmartyDNS, and VPNUK are all universal.  All  open proxies are also universal proxies, although the reverse does not hold for ibVPN and VPNUK.    Proxies that are open and universal  (CactusVPN, HideIPVPN, and SmartyDNS) allow any Internet user to proxy HTTPS traffic to any site, without having to register (or pay) for the \sdns service.

\Section{Information Leakage through DNS Probing} \label{sec:popularity}
\threatmodel{\sdns Provider}

Most \sdns providers advertise a number of channels (i.e., web sites) for which they will proxy traffic.  In this section, we describe how DNS cache probing techniques can be used to
\iflongversion
infer both (i)~the relative popularity of channels among a service's customers and (ii)~the number of users of an \sdns service.  Channel popularity allows us to gauge which sites' geofences are most often bypassed, while the number of users of an \sdns service enables us to estimate the revenue and general profitability of the service.
\fi
\sdns services do not publish statistics that describe which channels are actually
accessed by their customers, nor do they provide the 
 relative popularity of the channels that are accessed.  The DNS probing techniques described in this section provide a first glimpse as to how \sdns customers use these services.

\Subsection{Inferring Channel Requests}
To identify the channels requested by \sdns customers, we use the DNS cache snooping technique introduced by Grangeia~\cite{grangeia2004dns,dns:privacy:rfc} to determine whether or not the zone record for a hostname is in the cache of a resolver.

By default, clients almost always set the {\sc Recursion Desired} (RD) bit when sending queries to DNS servers.  This is true of many major operating systems  (including OSX, Windows, and Linux) and is intended to allow for the possibility of recursive DNS resolution.  In brief, the RD bit indicates to the resolver that the client prefers that the resolver perform a recursive DNS lookup.  Grangeia's cache snooping technique leverages the behavior that when the RD bit is {\em not} set, DNS servers (i)~respond with the resolved IP address if the entry is in its cache and (ii)~return either an error or the root name servers for the requested domain if it is not.  (We note that returning the root name servers is the expected behavior for iterative DNS resolutions.)  By setting the RD bit to zero, we  definitely learn whether the requested hostname is in the resolver's cache. 

\begin{figure}[t]
  \centering
  \begin{small}
  \begin{minipage}[t]{.30\linewidth}
    \centering
    \includegraphics[width=\linewidth]{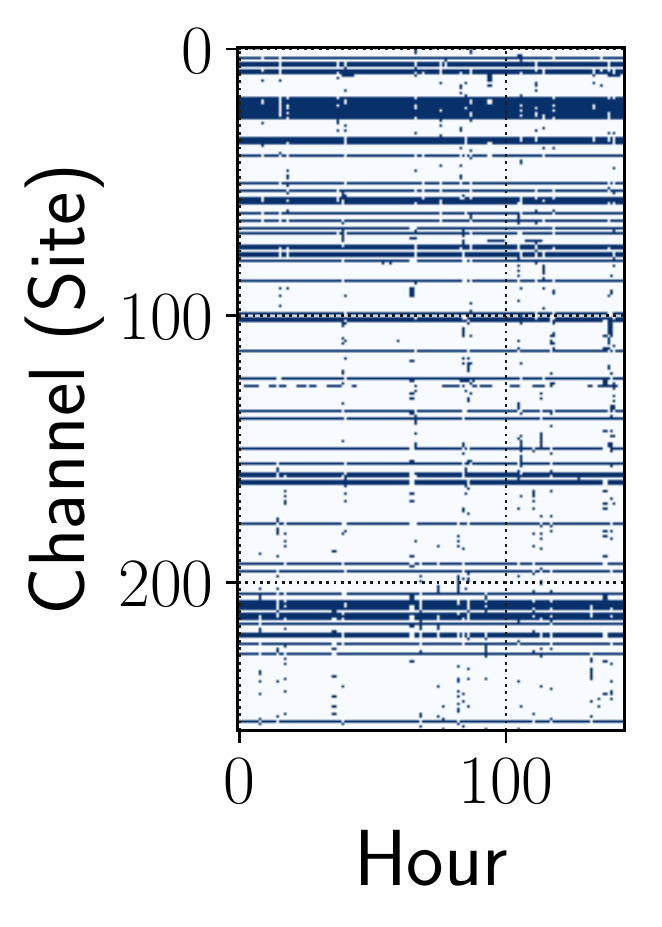}\\
    (a)~CactusVPN
  \end{minipage}
  \hfill
  \begin{minipage}[t]{.30\linewidth}
    \centering
    \includegraphics[width=\linewidth]{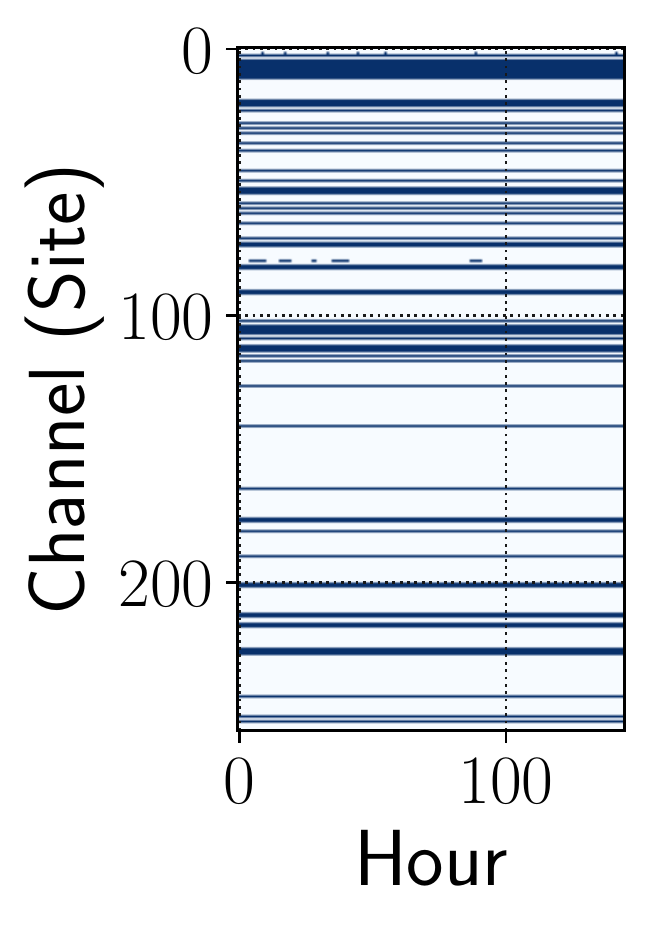}\\
    (b)~SmartyDNS
  \end{minipage}
\hfill
  \begin{minipage}[t]{.36\linewidth}
    \centering
    \includegraphics[width=\linewidth]{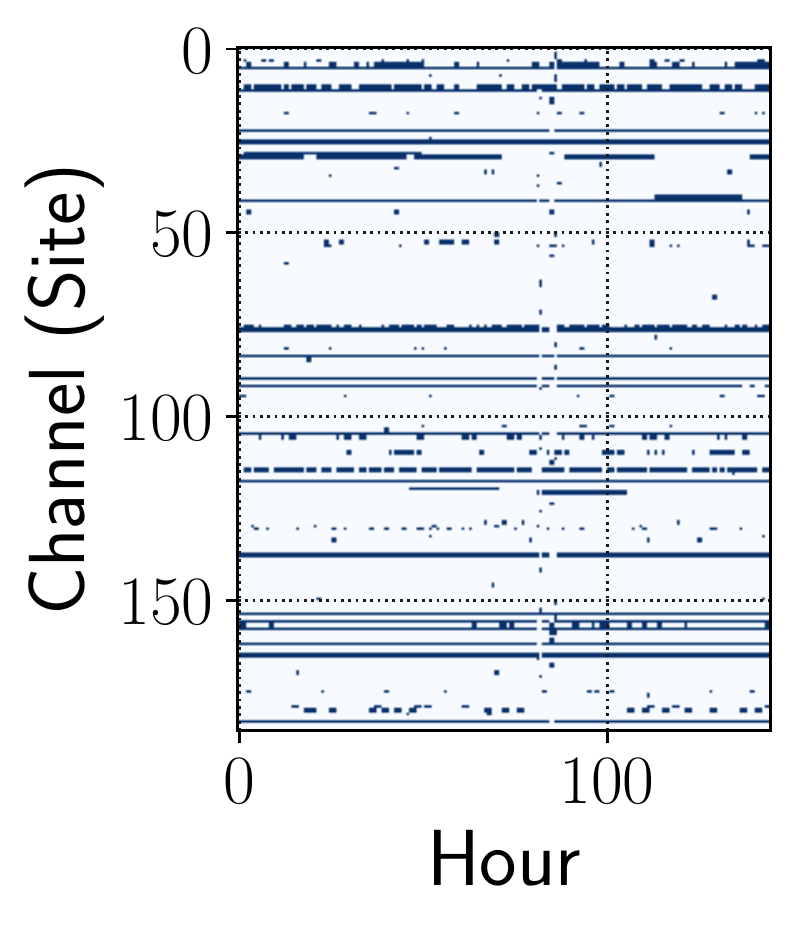}\\
    (c)~Unlocator
  \end{minipage}  
\end{small}
\caption{Presence (blue) and absence (white) of advertised channels' domain names in \sdns providers' DNS resolver caches.}
\label{fig:heatmaps}
\end{figure}

Figure~\ref{fig:heatmaps} shows the presence and absence of the hostnames for advertised channels on three \sdns providers over an approximately 5.25 day period beginning on August 21, 2019.  We probed each \sdns provider's cache once per hour during this period.   (We performed the experiment for other \sdns providers and obtained similar results; they are omitted for brevity.) Specifically, we examined the cache of the first DNS resolver that was listed on the webpages of the  three tested \sdns providers.  We observe that (i)~most of the domain names associated with the advertised channels never appeared in the resolvers' caches and (ii)~the few sites that did appear, did so consistently.  For example, less than 24\% (61 out of 256) of the sites supported by the SmartyDNS proxy ever appeared in its cache; TrickByte had the highest cache saturation with 61\% (50 out of 82) of its supported channels appearing at least once in its cache during our measurement period.  In summary, while \sdns providers advertise support for a large number of channels, our findings suggest that customers regularly use only a modest fraction of those offerings.

\newcommand{\ttl}{\ensuremath{\textrm{TTL}_l}\xspace}
\newcommand{\ttlmax}{\ensuremath{\textrm{TTL}_\textrm{max}}\xspace}

\tabpopularity{}

\Subsection{Deriving Channel Popularity}
Based on our prior analysis, we sought to understand the relative popularity of channels provided by \sdns.
We estimate  how often an \sdns provider's customers request each of the provider's supported channels by assuming the rate at which the \sdns provider resolves requests for a particular hostname is indicative of frequency at which its customers request it.

To perform this analysis, we use the popularity inference  technique of Rajab et al.~\cite{rajab2010peeking}, which operates as follows: a request for resolving hostname $H$ is sent to a DNS resolver $D$.  In its reply, $D$ returns the time (\ttl) at which the entry for $H$ will be expunged from its cache.  The maximum possible TTL (\ttlmax) can be obtained by querying the authoritative name server for $H$.  Rajab et al.'s technique issues a probe for $H$ once per \ttlmax, which allows for computing the {\em refresh} time (the time at which the cache entry for $H$ was most recently refreshed) as $T_r = T_p - (\ttlmax - T_l)$ where $T_p$ is the time of the probe.  This allows for the computation of the average rate $\lambda$ at which $H$ is requested from $D$ as
\[
  \lambda \approx \frac{R}{\sum_{i=1}^R(T_{r_i} - T_{r_{i-1}}-TTL)}
\]
where $R$ is the number of probes for $H$~\cite{rajab2010peeking}.

We implemented Rajab et al.'s algorithm and deployed it on 11 resolvers belonging to 11 different \sdns providers (i.e., we used one resolver per \sdns provider).  Three of the 11 resolvers reported erratic or otherwise erroneous TTLs, and we excluded these resolvers from our analysis.  For the remaining eight \sdns services, the average aggregate request rates, measured in requests per hour, are listed in Table~\ref{tbl:channelpopularity}.   We report the 10 most frequently requested hostnames for each \sdns service.  We additionally compute 95\% confidence intervals of $\lambda$ by applying the central limit theorem~\cite{rajab2010peeking}; this allows us to gauge our confidence in the results based, in part, on the number of probes we performed.  (Hostnames that have a large \ttlmax value resulted in fewer probes.)

Our findings reveal that streaming video---the target offering for many of the \sdns providers---is by far the most common destination for \sdns customers.  Interestingly, \sdns providers commonly resolve queries for popular news  (cnn.com) and social media (instagram.com) sites.  This suggests that \sdns customers regularly use \sdns resolvers and do not reserve their use for accessing geofenced content.

In Appendix~\ref{sec:estimates}, we use similar techniques to estimate the number of customers and revenue for several \sdns providers.

\Paragraph{Limitations to Popularity Measures}  To minimize the volume of our requests, we target only one DNS resolver for each \sdns provider.  Probing the unexamined DNS resolvers may yield different results.  Additionally, as shown in Table~\ref{tbl:channelpopularity}, the confidence intervals can be large, sometimes overwhelming a site's estimated arrival rate ($\lambda$). In such cases,  these results should be viewed with some skepticism.  Finally, our results assume DNS servers correctly follow the DNS protocol as described in the RFC~\cite{dns:rfc}.  Although, by definition, \sdns resolvers do not always return correct DNS responses, we observe that the tested \sdns resolvers appear to  generally adhere to the DNS RFC, with the sole exception of  returning false IP addresses for proxied channels.

 \Section{Discussion}

Some of the attacks identified in this paper are inherent to the
design of \sdns systems, and are difficult to remedy.  In particular,
the real-time \sdns user identification attack
(\S\ref{sec:deproxying})---which can also identify \sdns proxy
servers---is effective because \sdns can {\em only} proxy traffic that
first requires a DNS resolution.  Fundamentally, the attack exploits
the defining characteristic of \sdns services (i.e., the assignment of
proxies via DNS resolution), and thus we believe it is unlikely that
an \sdns provider can counter this inherent weakness.  On the other hand, the client enumeration attack
(\S\ref{sec:enumeration}) exploits an implementation flaw in some
\sdns systems, and can be remedied; in fact, one provider
implemented a fix after we disclosed the attack.

The increased risk of traffic analysis (\S\ref{sec:eavesdropping}) is
similar to the risk that arises from using VPN services: longer Internet paths
generally provide more opportunities for eavesdropping.  However,
unlike with VPNs, \sdns has no easy on/off switch, and we suspect that
many \sdns users configure their computers to use \sdns resolvers and
do not restore their settings after using \sdns.  This ``set and
forget'' functionality is unusual for VPNs, which typically require
user interaction to enable.  We conjecture that \sdns use therefore
provides a more persistent level of susceptibility to eavesdropping
than VPNs due to the likely longevity of using \sdns resolvers.

Encrypting DNS traffic between a client and the \sdns resolver using
either DoH~\cite{doh:rfc} or DoT~\cite{dot:rfc} mitigates some of the
eavesdropping risks.  However, encrypted DNS is still subject to
traffic analysis which can leak the sites being
resolved~\cite{bushart2020padding,siby2020encrypted,houser2019investigation}
to on-path  eavesdroppers.
Perhaps more importantly, not all major operating systems support DoH or DoT.  We
believe at least for the short-term that it is unlikely that \sdns
providers would add support for DoH/DoT, since doing so may increase
the level of technical sophistication required to configure \sdns.

The authentication and authorization failures (\S\ref{sec:authfail})
leverage poor design decisions by the vulnerable \sdns services.
Applying IP-based authentication at both the resolver and the proxy
prevent unauthorized use.  (It is worth emphasizing that IP-based
authentication does not provide strong authentication.)  

Finally, the exposure to analytics (\S\ref{sec:popularity}) uses DNS
cache probing techniques that are generally applicable to DNS
resolvers.  They are arguably especially problematic however in the
\sdns setting since the sole purpose of \sdns services is to bypass
geofencing, and hence determining how often these services are used
for each channel could be useful to assess the potential criminal
culpability or legal liability of the providers.
To prevent such information leakage, DNS resolvers (whether
\sdns resolvers or ordinary resolvers) could advertise the maximum TTL
value, although such behavior would clearly violate the DNS
specification~\cite{dns:rfc}.
 \Section{Ethical Considerations}
\label{sec:ethics}

At all times, we sought to minimize risk, both to the users of \sdns services and the services themselves.  Our experiments were guided by the principles outlined in the Menlo Report~\cite{menlo-report}. We use this guideline to elaborate on the ethics of our study:

\Paragraph{Respect for persons}
We avoided causing harm to individual users through our measurements and proof-of-concept attacks by not targeting specific individuals (other than ourselves). In validating the client enumeration attack in \S\ref{sec:enumeration}, we used a small-scale proof-of-concept in which we spoofed only our own IP addresses. We
did not attempt to discern whether any IP addresses, other than those
operated by the authors, belonged to customers of the vulnerable \sdns
services.  We did not issue more than a handful of queries to the
\sdns provider's resolver, and our DNS queries were all well-formed and
conformed to the DNS standard~\cite{dns:rfc}.

Similarly, in \S\ref{sec:popularity}, when identifying \sdns usage rates and the popularity of channels, our measurements consisted of sending a relatively low volume (one request per hostname per the hostname's \ttlmax) of well-formed DNS queries to a DNS resolver.  This volume of requests is negligible compared to the request arrival rate of \sdns providers (see Table~\ref{tbl:channelpopularity}).  This measurement provides respect for individuals and persons in that it does not directly interfere with the normal activities of the \sdns service nor its customers. In brief, we used \sdns resolvers exactly as they were intended to function, and merely inspected the returned result (specifically, its TTL value) to derive statistical inferences.

\Paragraph{Beneficence}
At all stages of our study, we sought to reduce harm and maximize the benefits of the research. Foremost, we designed our experiments to avoid overwhelming the public DNS or \sdns resolvers by rate limiting our requests, and we only submitted well formed DNS, HTTP, and HTTPS requests throughout. 

Most relevant, after identifying the client enumeration attack, we performed responsible disclosure and notified the operators of the two affected services (VPNUK and ibVPN).We received a response from ibVPN the following day, informing us that they had (partially) mitigated the attack (see~\S\ref{sec:enumeration}).  The goal of disclosing the attack was to decrease the associated risks identified and increase the overall benefits of the research. 

\Paragraph{Justice}
Our measurements are just in that we spread out all probes to a broad set of \sdns providers, treating each equally in our experimentation without targeting specific services over others.

\Paragraph{Respect for Law and Public Interest}
We designed our study to be both lawful and in the public interest. First, we paid or used free-trials for all of the measured \sdns providers; we did not use these services surreptitiously.  Additionally, we considered and reported on the privacy and security ramifications of using these services, which is in the public interest, and when we identified potential vulnerabilities, we responsibly disclosed them to the \sdns providers. In taking this action, we consulted with our institution's general counsel to ensure that we took action in a legally responsible manner (based on our jurisdiction).

 \Section{Conclusion}

This paper presents the first study of smart DNS (\sdns) services in the wild. 
  We identify a number of architectural weaknesses in currently deployed \sdns services that affect the privacy and security of end users, who may be confused with the privacy properties of \sdns as compared to other services that can avoid geographic restrictions, such as VPNs.
We show that content providers can trivially detect \sdns users who access their sites, identify these users' unobscured IP addresses, and enumerate {\em all} customers who are registered for these \sdns services (whether they access the content provider's website or not). Worse, some settings of \sdns allow {\em any} arbitrary third party to enumerate all customers of an \sdns service, even when those users are offline. These vulnerabilities were confirmed and repaired by an \sdns provider after responsible disclosure.

We also identify a number of authentication and authorization errors in the setup of \sdns, with many services  relying only on IP-based authentication at the DNS server and neglecting to perform these checks at the proxy server. The failure to properly authenticate and authorize users effectively transforms \sdns providers' proxies into a distributed network of open SNI proxy servers. We describe a straightforward method of discovering these open proxies and discuss how unscrupulous users could use \sdns services while bypassing payment. Additionally,  we show that some \sdns providers proxy content that is not advertised as being supported, raising  further privacy concerns about  traffic interception and manipulation.

Although  \sdns  provides  customers with the ability to  seamlessly bypass geofenced content, this comes at the expense of a large number of privacy and security risks, including the exposure of \sdns service usage and individual browsing habits.

\section*{Acknowledgments}

We are grateful to the anonymous reviewers for their insightful feedback, and especially to our shepherd, Tariq Elahi, for the many suggestions that helped us improve this paper.  We also would like to thank Tavish Vaidya for the many fruitful conversations about \sdns services.

This paper is partially supported by the National Science Foundation under grants \#1718498 and \#1845300.  The opinions, findings, and conclusions or recommendations expressed in this paper are strictly those of the authors and do not necessarily reflect the official policy or position of any employer or funding agency.
 
\bibliographystyle{plain}

\appendix

\setcounter{table}{0}
\renewcommand{\thetable}{\thesection.\arabic{table}}
\setcounter{figure}{0}
\renewcommand{\thefigure}{\thesection.\arabic{figure}}

\section{Additional Methodological Details and Ecological Findings}
\label{sec:measure}

To bypass geofencing restrictions, \sdns services rely on a network of strategically placed proxies and DNS resolvers. As part of our analysis and to better understand these services’ ecosystem, we attempted to answer the questions: \emph{where are \sdns resolvers and proxies located?; who hosts them?; and what was the likely motivation behind these choices?}

As a first step, to better understand to whom the \sdns providers are catering their services, we determine the geographic location of the \sdns providers' DNS resolvers.

To reduce network latencies incurred during DNS resolutions, \sdns providers often recommend that their customers select a \sdns resolver that is in close physical proximity.  Understanding where \sdns providers place their resolvers thus provides hints as to where they envision the best opportunities for attracting customers.  Using MaxMind's geolocation service, we map the listed DNS resolvers for each proxy to a location, and report the countries with the most resolvers in Table~\ref{tab:resolver:loc}.  With the exception of the United States, we note that the locations of potential customers mostly differ considerably from the locations in which the \sdns providers operate (see Table~\ref{tab:providers}).  In short, it appears that \sdns providers mostly tailor their services to international customers.

While the locations of the resolvers indicate potential customer markets, the proxies' locations correlate to the geo\-fences that the \sdns providers aim to bypass.  
However, due to the nature of the modern web and the prevalence of content distribution networks (CDNs), identifying \sdns providers' proxy servers proved  complicated.

At a high level, the task entails querying an \sdns resolver for a hostname and identifying whether the returned IP address was (i)~accurate or (ii)~that of a proxy server. In reality, unfortunately, DNS resolution is fairly complex.  Rather than consistently returning a single IP, multiple queries to a single domain name on a normal (non-\sdns) resolver return the IP of the host that can serve the website's content fastest, given the current network state and the requester's network location. When accounting for the widespread use of CDNs, this also means seemingly unrelated sites can be resolved to the same IP address (since multiple sites can share CDN replicas).  In short, it is difficult to enumerate all possible valid IP addresses that belong to a given hostname; this is especially true of popular sites (including the channels supported by \sdns providers), since such sites tend to heavily rely on CDN services.

We identify  proxy IPs using a two-phase approach: at a high level, we first identify a set of {\em candidate IPs} we believe may be \sdns proxies, and then verify them.
To generate candidate IPs, we queried 10 \sdns providers' resolvers for two sets of domains: the Alexa top 1,000 most popular sites~\cite{Alexa-top1m:online}, and the hostnames of the channels advertised as being supported by the \sdns provider. 
We then compared the returned 
list of IPs against a {\em ground truth} dataset generated by making over 32,000 DNS requests to Google's and CloudFlare's DNS resolvers, as well as requests from RIPE Atlas probes to their local resolvers.  The latter was included to increase the geographic and network diversity of the requesting DNS clients.  The ground truth dataset was constructed using DNS queries conducted between February~14 and  March~31, 2019, and again between April~25 and May~3, 2019.
Finally, we generated a  candidate list of hostname-to-resolved-IP pairings by first considering the responses from \sdns resolvers and then eliminating entries for which an IP in the same /24  appeared in the ground truth dataset.

To verify the candidate IPs as proxies, we attempted to fetch content via a candidate proxy from both a machine that was registered with the \sdns service and one that was not.  Conceptually, if the candidate IP is not a proxy and is a legitimate IP address that serves content for the site, it should serve the content regardless of the requestor's IP; on the other hand, if it {\em is} a proxy, then the proxy should only serve content for the IP that is associated with one of its customers.  That is, we expect actual proxies to serve requests that originate from a registered IP, but to deny the same content requests from IPs that are not associated with the \sdns provider. 

Using two machines, one whose IP we registered with the \sdns service,  and one whose IP was not registered, we sent well-formed HTTP/S requests (with the {\sc Host} HTTP and SNI headers properly set) to the candidate proxy using \htmlize{curl}, and compared the results.  We confirm a candidate IP as an actual proxy if and only if the HTTP/S request from  the registered machine was successful (i.e., resulted in a {\sc 200 OK} HTTP response) and the request from the non-registered machine was not.

Overall, we were able to definitively identify 54 distinct proxy IPs across five of the evaluated \sdns providers.

Table~\ref{tab:proxy:loc} lists the most common countries where proxy servers are located.  We note that the most popular locations---the United States, the United Kingdom, and India---are also the nations that host a large fraction of the channels offered by \sdns providers.  This suggests that proxies are indeed placed close to content providers.

\tablecounts{}

 \section{Mapping an IP Address to an Individual}
\label{sec:ip2person}

Although mapping an IP address to an autonomous system (AS) is straightforward, the process of matching an IP to an actual individual is more difficult, error-prone, and less well-understood.
This is due to a number of different factors including (but not limited to) the widespread use of DHCP for IP address allocation, and ISPs' widely varying policies for managing pools of available IP addresses. The former is especially problematic, since DHCP does not set any explicit requirements for how long each IP address is allocated to a single entity (e.g., a household)~\cite{dhcp:rfc}.  Nor does DHCP require a lessee to notify the DHCP server if it relinquishes an IP address before its lease expires~\cite{dhcp:rfc}.

In the context of the IP enumeration attack described in~\S\ref{sec:enumeration}, for each IP address found to be registered with an \sdns provider, the attacker can determine the ISP, AS, and rough geographic location (with the caveat that IP geolocation services are not always accurate) from which the IP address originates.

However, the attacker can use additional datasources to sometimes hone in on the household or even individual who leases a particular IP address. Many companies collect publicly available records and mine information from sources such as social media, web beacons, browsing histories, and user cookies placed across the Internet to create ``people databases''---vast databases that contain detailed profiles of Internet users.  These profiles often include an Internet user's full name, address, gender, age, phone number(s), email(s), date of the profile's last update, and---most relevant to our attacks---the user's IP address~\cite{thatsthem}.

Using these people databases as a primary backend, companies such as ThatsThem~\cite{thatsthem} offer search tools that allow anyone to map an IP address to an individual or household.  We note that these search tools do not have complete data, and it is difficult to assess their accuracy.  However, traditional and far more well-established data brokers, such as Experian, also offer IP-to-individual mapping services~\cite{experian}.  

It is worth noting that, in addition to their public facing offerings, consumer data collection companies frequently buy and sell collected information from/to each other, meaning that once data about an individual is collected by one company, that information likely propagates to others~\cite{experian, thatsthem:privacy}.

In summary, an adversary who learns an \sdns user's IP address could use these people databases to learn not only the identity of the \sdns customer, but also additional information (e.g., address and email).  For this reason IP addresses are sometimes considered personally identifying information under both GDPR and the California Consumer Privacy Act of 2018~\cite{gdpr, ccpa}, and that the Privacy Commissioner of Canada issued a report detailing the privacy risks of exposing IP addresses~\cite{canada-privacy}.

 \section{Estimating the Number of Customers and Revenue}
\label{sec:estimates}
Again borrowing from the technique of Rajab et al.~\cite{rajab2010peeking}, we can use the average request rate ($\lambda$) to form a rough approximation of the number of users ($n$) of an \sdns provider.  We denote $\lambda(\textrm{site})$ as the value of the aggregated (total) average request rate for a given site.  Further, let 
$\lambda_c(\textrm{site})$ be the expected request rate for a {\em client} accessing the site.  That is, while $\lambda(\textrm{site})$ denotes the total  requests per unit time for the site, $\lambda_c(\textrm{site})$ is the number of requests due to a given user.  Then, assuming Gamma distributed arrival times, Rajab et al.~shows that the number of users $n$ of an \sdns provider is:
\[
  n = \frac{\lambda(\textrm{site})}{\lambda_c(\textrm{site})}
\]

Rajab et al.~reports that $\lambda_c(\textrm{google.com})$ is 2.63 requests per hour~\cite{rajab2010peeking}.  We can compute $\lambda(\textrm{google.com})$ for the various \sdns providers using the technique described in the previous section, and thus compute $n$. Here, we perform our probes (i.e., DNS requests for google.com) over an approximately \confirmed{11} hour period beginning on September~6, 2019.
To limit the load on the \sdns providers' resolvers, we send  probes to a single resolver per \sdns provider.

Table~\ref{tbl:estusers} lists the empirically measured average request rate for google.com and the derived number of users for six \sdns services' resolvers. (The remaining providers did not  consistently respond properly to DNS requests to resolve google.com, and are excluded from the Table.)  Note that the number of estimated users in Table~\ref{tbl:estusers} is based on traffic to a single resolver (per service) and thus likely undercounts the total number of users of a service.

Of the successfully tested providers, we find that CactusVPN has approximately 16K users using a single one of its resolvers, while the other \sdns services have significantly fewer users accessing their tested resolvers.

Using the pricing information presented in Table~\ref{tab:providers}, we can then estimate the revenue for each \sdns provider by multiplying the estimated number of users by the price-per-user.
This should be considered a conservative (low) estimate of the provider's revenue, since our probing DNS requests target only the first listed DNS resolver for each \sdns provider. 

We can estimate profit margins for an \sdns provider based on the expected costs of running proxy servers.  Proxies relay content to/from supported channels, and we consider a near-worst case scenario in which all \sdns users continuously stream high-quality video.  Here, we use Netflix's reported bandwidth requirements of 3~GB/hour (6.67~Mbps) to estimate  \sdns providers' bandwidth needs.  \sdns providers can easily support such rates with VPS providers.  In particular, there are a number of VPS providers that provide uncapped (sometimes called {\em unmetered}) 1~Gbps links~\cite{uncensoredhosting} for approximately \$ 10 per month.  We note that a single 1~Gbps link can support 150 \sdns customers (each of whom consumes 6.67~Mbps).  The revenue from 150 \sdns customers far exceeds the bandwidth costs.  For example, CactusVPN charges \$\,4.99 per customer, per month, for a revenue of \$\,748.50 and profit of \$\,738.50 (after subtracting the \$\,10 VPS cost) per 150 customers.  The profit per customer is thus \$\,4.92 per month, yielding a profit margin as high as 98.6\%. Using these general assumptions, we provide estimates of the profits of \sdns providers in Table~\ref{tbl:estusers}.

\tabrevenue{}

\Paragraph{Limitations to Profit and Revenue Estimation}  Our analysis relies on a number of assumptions, including the expected distributed arrival times and the accuracy of  $\lambda_c(\textrm{google.com})$ reported by Rajab et al.~in 2010.  We note that the client enumeration attack presented in \S\ref{sec:security} constitutes a far more accurate method of determining the precise number of \sdns customers, although the attack only works for a subset of \sdns providers. (Due to obvious ethical concerns, we did not perform the enumeration attack described in \S\ref{sec:security} to measure \sdns usage.) 

Additionally, as discussed above, a more complete revenue exploration of an \sdns service would include the infrastructure costs of resolvers, as well as the fact that not all proxies are fully utilized. Further, our analysis ignores the (potentially high) costs of customer support and maintaining an infrastructure for billing.

\iflongversion
  \section{Performance of \sdns Services}
\label{sec:performance}

DNS resolution is a frequent operation not only for web browsing, but
also for myriad other applications and system services.  For \sdns
users, DNS resolution requests are handled by the \sdns resolver
(absent local caching, which is rare).  Relative to using a local
ISP-provided resolver, 
 resolving hostnames through \sdns imposes significant
performance costs: local DNS resolvers are
typically located near the requesting client~\cite{mao2002precise},
producing low roundtrip times.  In contrast, \sdns providers have
scant offerings of DNS resolvers from which to choose, making it
unlikely that the chosen resolver will be close to the client.  Second,
due to the prevalence of CDNs, the mapping from hostnames to IPs may be
dependent upon the location of the client IP and the DNS server.  \sdns resolvers
serve diverse clients and hence its cached entries (and, consequently,
its returned IPs) are less catered to any particular client's network
location.

We empirically measure the cost of using \sdns resolvers by performing
DNS lookups to the top 1000 Alexa sites~\cite{Alexa-top1m:online} via
(i)~\sdns resolvers, (ii)~Google's free DNS resolver at \htmlize{8.8.8.8}, and (iii)~our local
institution's default resolver.  We note that Google's public DNS service uses
IP anycast to ensure that resolution requests are routed to the
closest resolver~\cite{anycast8888}.
For each resolver and Alexa hostname,
we performed five DNS resolutions (without caching) from a client
machine on our institution's wired network.

We find that, unsurprisingly, 
the local resolver has the lowest median response time, closely
followed by Google's DNS server. Using local
resolution or Google's highly replicated DNS server offers orders of
magnitude better response times than any of the tested \sdns
providers.  For example, the HideIP resolver impose 
approximately a 2500\% overhead in median response time relative to
our local resolver.

We also consider how the use of \sdns services affects overall
browsing experience, as measured using web page load
times.  Here, the effects of slow DNS resolutions could be compound by
the number of web objects embedded on a webpage, since objects at
difference domains (or subdomains) require DNS resolution.

To examine the effects of \sdns usage on web page load times, we instrumented Selenium using
the Chrome 76.0.3809.126 driver for Linux and configured it to use one
of the three 
DNS configurations:  our local institution's
resolver, Google's 
open resolver, and  the \sdns resolver.  Note
that for channels supported by the \sdns provider, using the 
provider's DNS resolver also meant that at least some web content will
be relayed via one or more of the provider's proxy servers, since
Chrome will fetch content from whatever IP addresses are resolved via
DNS.  We include Google's resolver to consider cases in which
resolution is not performed locally, but the returned results are
correct (i.e., the content is not proxied).  In all cases, the
headless Chrome browser fully renders the requested page.

\begin{figure}[t]
  \centering
  \begin{minipage}[t]{1.0\linewidth}
    \begin{footnotesize}
      \centering
      \includegraphics[width=\linewidth]{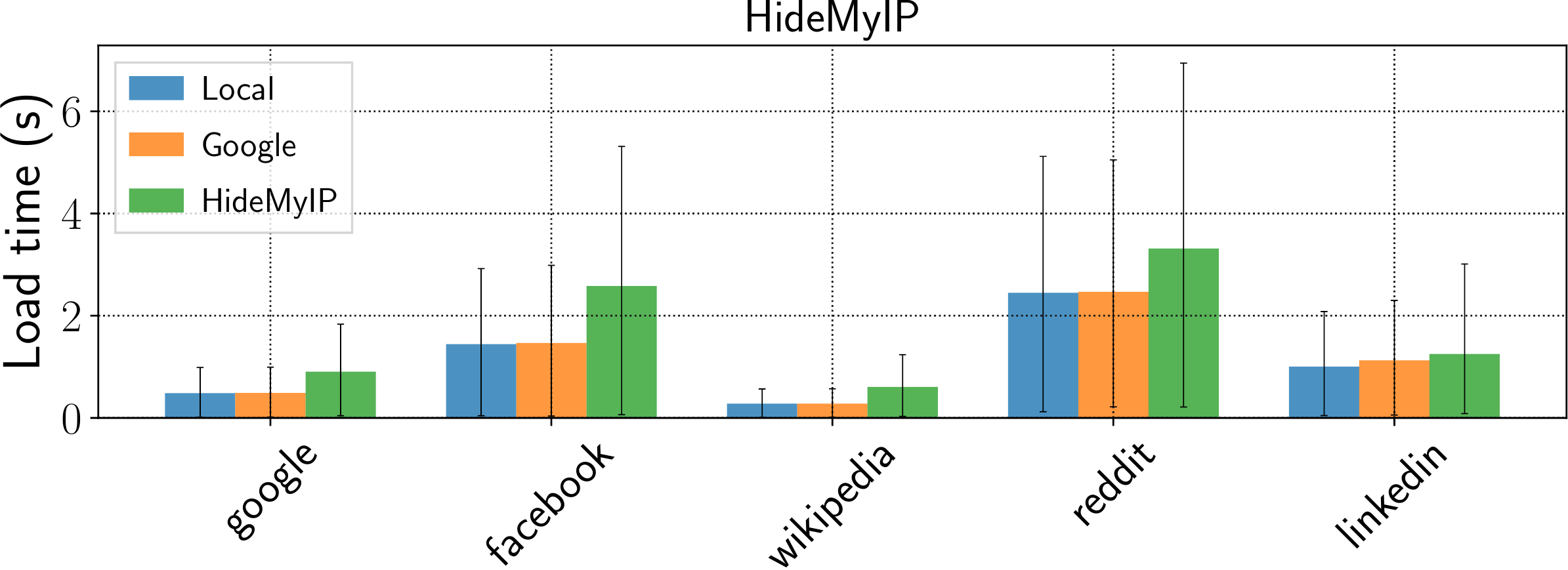}
    \end{footnotesize}
  \end{minipage} \\
  \begin{minipage}[t]{1.0\linewidth}
    \begin{footnotesize}    
      \centering
      \includegraphics[width=\linewidth]{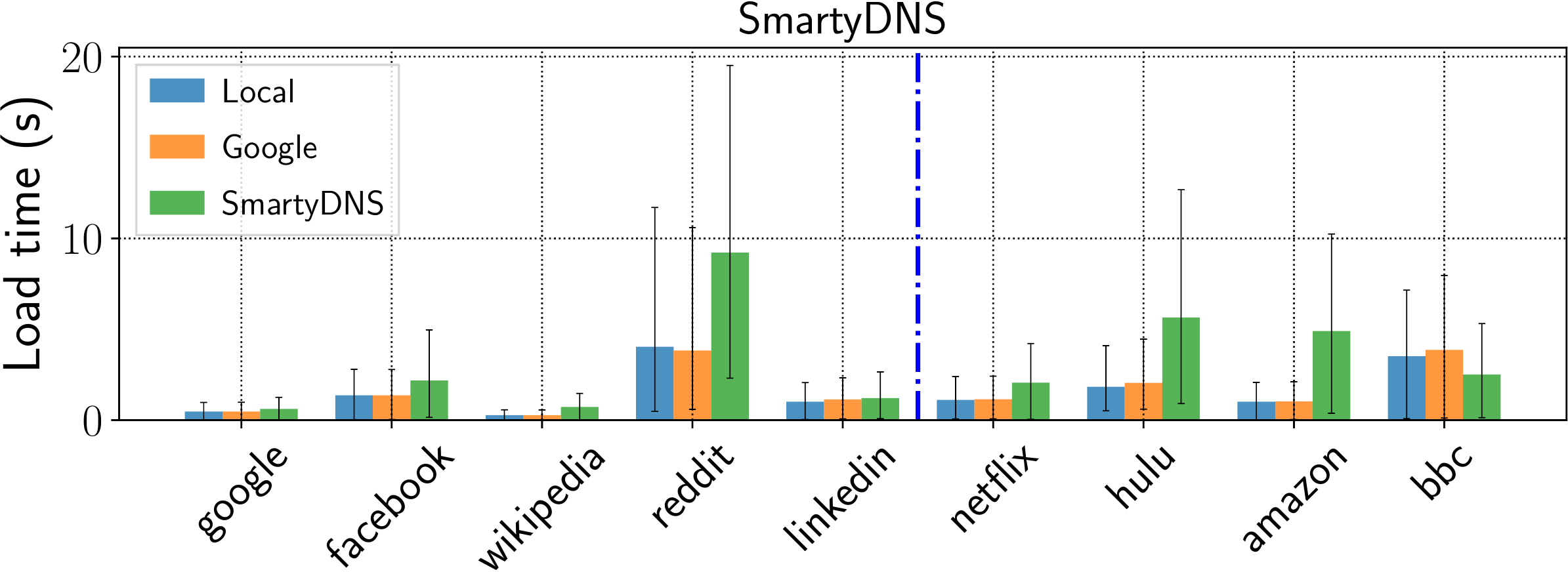}
    \end{footnotesize}
  \end{minipage}
\caption{Page load times for top pages of various sites, when using
    the local DNS server, Google's open DNS server at 8.8.8.8, and the
    \sdns DNS resolver to resolve hostnames.  Error bars show the
    interquartile range.  The vertical dashed blue bar for SmartyDNS
    and CactusVPN separate sites that are not advertised as being
    proxied (left of the line) and those for which the \sdns
    advertises support.}
  \label{fig:perf_e2e}
\end{figure}

Figure~\ref{fig:perf_e2e} shows the median page load times for various
websites, over \confirmed{50} fetches performed on
\confirmed{September 7th, 2019}.  Error bars indicate the
interquartile ranges.  We focus on two particular \sdns providers for
brevity; we obtained similar results for other providers (not shown).  We include both sites that are proxied by the
\sdns provider (right of the dashed horizontal bar) and those that are not
proxied (left of the horizontal bar).  The latter is included since we
posit that most \sdns users will not regularly manually update their
DNS settings and will instead rely consistently on their \sdns
provider's resolver.  As a result, the performance of any
communication that depends on DNS resolution will be affected by the
latencies incurred by using a remote resolver and potentially due to
over-proxying (see~\S\ref{sec:overproxying}).

The HideMyIP service, whose performance is shown at the top of
Figure~\ref{fig:perf_e2e}, is a unique case because it proxies all
network communication.  Its DNS resolver always returns its own IP;
HTTP/S traffic sent by the client is always proxied through this
combined resolver/proxy.\footnote{We note that configuring a computer
  to use HideMyIP (i.e., by changing its network settings) prevents
  the use of non-HTTP-based protocols.  HideMyIP is able to proxy all
  HTTP/S traffic by inspecting the {\sc Host} HTTP header or the TLS
  SNI header (see~\S\ref{sec:architecture}).  However, repeating the
  requested hostname in the application-layer payload is generally a
  rarity in network protocols.  The HideMyIP proxy is unable to
  determine the actual requested destination when other protocols that
  do not explicitly include the domain name are used (e.g., ssh), and
  therefore cannot proxy these protocols.  Since the DNS resolver is a
  machine-wide setting, the use of HideMyIP significantly restricts
  the types of communications the computer can use.} For the tested destinations, HideMyIP imposes fairly
substantial overhead.  For example, it increases the median load times
by \confirmed{79\%} and \confirmed{76\%} for rendering Facebook's top
page, when compared against retrieval when  local DNS and Google DNS
resolutions are performed, respectively.

The increases in webpage rendering time were also pronounced for the
SmartyDNS service, shown in the bottom plot of
Figure~\ref{fig:perf_e2e}.  For the non-proxied sites (that appear to
the left of the dashed blue line), the differences in performance when
using local versus Google resolution were minor.  However, SmartyDNS
incurs fairly significant overheads, especially in the case of reddit.
We note that the top page of reddit requires the retrieval of more
than 180 embedded web objects.  The overhead of using SmartyDNS to
resolve these requests is significant, resulting in a
\confirmed{128\%} increase in load page time over using the local resolver.
Using SmartyDNS also results in longer page load times for sites that
are advertised as being supported (i.e., proxied) by SmartyDNS.  As an
interesting outlier, retrieving the webpage bbc.co.uk was {\em faster}
when using SmartyDNS.  We suspect that this may be due to either a
triangle-inequality violation in the Internet
topology~\cite{tiv-aware} in which routing through the proxy yields a
higher performing connection than the default route, or (perhaps more
likely) cacheing that is performed on the proxy node.

 \fi
\end{document}